\documentclass[twocolumn,showpacs,pre,aps,longbibliography,superscriptaddress]{revtex4-1}
\usepackage[bookmarks=false,linkcolor=blue,urlcolor=blue,colorlinks,citecolor=blue]{hyperref}
\pdfoutput=1 

\usepackage{amsmath}
\usepackage{amssymb}
\usepackage{graphicx}
\usepackage[shortlabels]{enumitem}
\usepackage[retainorgcmds]{IEEEtrantools}
\usepackage[bottom]{footmisc}
\usepackage{caption}
\usepackage{subcaption}

\begin{document}

\title{Relevance in the Renormalization Group and in Information Theory}
\author{Amit Gordon}
	\affiliation{Racah Institute of Physics, The Hebrew University of Jerusalem, Jerusalem 9190401, Israel}
\author{Aditya Banerjee}
	\affiliation{Racah Institute of Physics, The Hebrew University of Jerusalem, Jerusalem 9190401, Israel}
\author{Maciej Koch-Janusz}
	\affiliation{Department of Physics, University of Zurich, 8057 Zurich, Switzerland}
	\affiliation{James Franck Institute, The University of Chicago, Chicago, Illinois 60637, USA}
\author{Zohar Ringel}
	\affiliation{Racah Institute of Physics, The Hebrew University of Jerusalem, Jerusalem 9190401, Israel}

\begin{abstract}
	The analysis of complex physical systems hinges on the ability to extract the relevant degrees of freedom from among the many others. Though much hope is placed in machine learning, it also brings challenges, chief of which is interpretability. It is often unclear what relation, if any, the architecture- and training-dependent learned ``relevant" features bear to standard objects of physical theory.	
	Here we report on theoretical results which may help to systematically address this issue: we establish equivalence between the information-theoretic notion of relevance defined in the Information Bottleneck (IB) formalism of compression theory, and the field-theoretic relevance of the Renormalization Group. We show analytically that for statistical physical systems described by a field theory the ``relevant" degrees of freedom found using IB compression indeed correspond to operators with the lowest scaling dimensions. We confirm our field theoretic predictions numerically. We study dependence of the IB solutions on the physical symmetries of the data. Our findings provide a dictionary connecting two distinct theoretical toolboxes, and an example of constructively incorporating physical interpretability in applications of deep learning in physics. 
\end{abstract}

\maketitle

The study of theoretical models is an essential part of physics. For sufficiently complex systems, however, establishing what the correct degrees of freedom are, and building a model in their terms, is a challenge in itself. The process is driven by experimental or numerical observations, but in practice physical intuition and prior knowledge are crucial to constructing a sufficiently simple model capturing the ``essence" of the phenomenon, rather than abundance of raw data \cite{doi:10.1098/rsta.2016.0153}. Still, data itself should contain sufficient information for this task, and a tantalizing prospect is to perform it in an unbiased, automatic fashion using modern computational methods, particularly deep learning (DL) \cite{RevModPhys.91.045002}. A fundamental obstacle to this is the mismatch between the concepts of physics, largely formulated in the language of field theory, and the theory and engineering practice of DL, all but ensuring questions of interpretability \cite{Murdoch22071}. To bridge this divide a framework is required capable of expressing, and allowing for practical computation, of quantities on both sides. Information theory, deeply connected to physics and computer science  \cite{preskill,10.5555/971143,10.5555/1592967} is a natural candidate.

In its classical formulation information theory was intentionally agnostic to the contents of the information, focusing on its efficient transmission \cite{Shannon48}. Though often only part of the information is pertinent to the problem, defining a formal notion of ``relevance" in sufficient generality has proven difficult~\cite{BlauM19}. This was addressed in the seminal Information Bottleneck (IB) paper~\cite{infbottle1}: relevant information in a random variable was defined by correlations, or sharing information, with an auxiliary ``relevance" variable, providing an implicit filter (an example of such correlated pairs are full frequency decomposition of a recorded speeches, and their written transcripts). Compressing data to preserve the relevant part most efficiently was cast as a Lagrangian problem, for which efficient DL methods have recently been introduced~\cite{AlemiFD016}.

In physics, however, there already exists a fundamental and \emph{a priori} independent notion of relevance, based on the properties of the operators under scale transformations embodied in the celebrated renormalization group (RG) flow\cite{Wilson1974,Wilson1975,Fisher1998}. RG relevance is the most precise definition we possess
 of what it means for an observable to determine macroscopic physical properties of the system; it directly connects to the powerful formalism of conformal field theories (CFT)\cite{BELAVIN_first,BELAVIN1984333,PhysRevLett.52.1575,DiFrancesco:639405}, which revolutionized the understanding of critical phenomena \cite{Cardy:318508,doi:10.1142/0608,RevModPhys.91.015002}.

Here we show that these two notions, belonging to entirely different theoretical frameworks, are in fact equivalent in physical systems, \emph{i.e.}~the information about long-range properties ``relevant" in the information-theoretic sense is formally determined by the most ``relevant" operators in the sense of RG.
Information loss in the context of RG has been attracting interest since the observation of irreversibility of its flow \cite{Zamolodchikov:1986gt,gaite-oconnor,Casini_2007,Apenko2012,Machta604,Balasubramanian:2014bfa,Beny2015a,Beny2015b,Beny2018coarsegrained}; we introduce a formal connection to compression theory which is constructive, quantitative, and \emph{computable}. This allows us to verify our predictions numerically.
We prove that using the IB approach the most relevant operators can be extracted from the data, along with information about physical symmetries. This result is thus not only of theoretical, but also of practical importance.
It provides a 
route towards automating 
theoretical tasks \emph{e.g}~deriving Ginzburg-Landau effective descriptions, and detecting symmetries hidden or emergent, in a controlled and by construction interpretable way, using the toolbox of statistics and machine learning. To wit, while we focus on theoretical foundations, in a parallel work 
these results and recent DL advances \cite{belghazi2018mine,poole2019variational} are leveraged to construct an efficient algorithm, the real-space mutual information neural estimator (RSMI-NE)\cite{RSMI-NE}, extracting the physically most relevant degrees of freedom from much larger inputs, along the way characterizing spatial correlations, phase transitions and order parameters. We show here that RSMI is a limit of the IB problem, providing a theoretical underpinning for this promising numerical method.

Below we briefly review IB theory and its relation to the RSMI approach to real-space RG in the context of statistical mechanical systems described by a CFT. We then present the main result: an analytical solution to the IB equations at strong compression which provides an explicit dictionary between IB relevancy, RG-relevancy, and eigenvectors of the transfer matrix in any dimension. We compare these predictions with numerics, obtaining agreement to high precision. In addition we show how symmetries are manifested in the compressed/coarse-grained degrees of freedom. Supplemental Materials give technical details and background information.

\captionsetup[figure]{justification=raggedright}
\begin{figure}[tp]
	\centering
	\includegraphics[width=8.5cm,trim={0.5cm 12cm 0 7cm}]{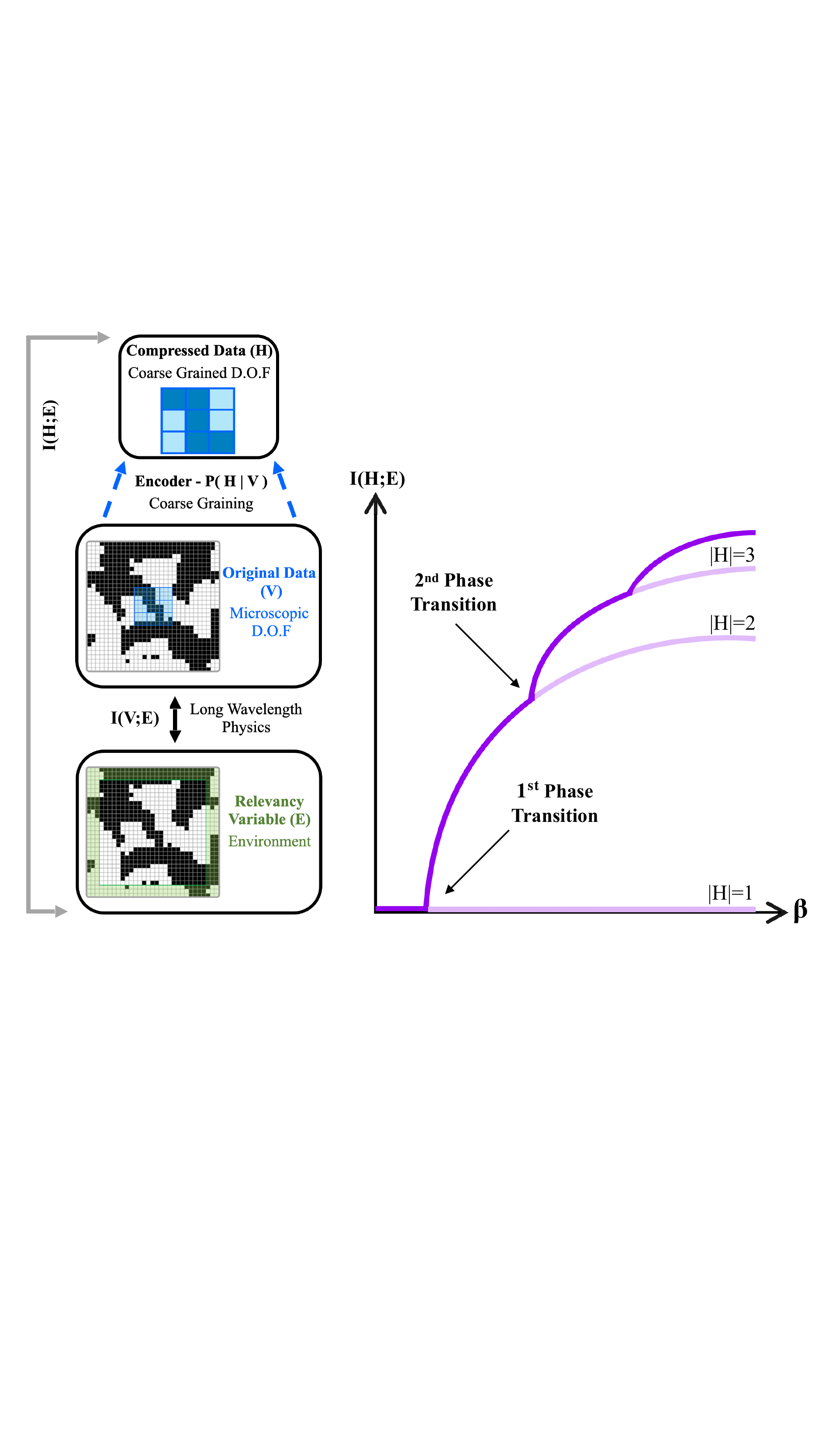}
	\caption[justification=justified]{\textbf{Left:} The general outline of the IB scheme, and in the physical setup of RSMI RG \cite{Koch-Janusz2018,PhysRevX.10.011037}: an optimal encoder extracting information about ``relevance" variable $E$ contained in $V$ is constructed. \textbf{Right:} IB curves depicting relevant information $I(H;E)$ retained by solutions to the IB equations (encoders), as a function of the tradeoff $\beta$ (see Eq.\ref{eq:main_ib_lag1}). At critical values of $\beta$ phase transitions occur: new solutions, with compressed variable $H$ of increased cardinality (\emph{i.e.}~tracking additional features) appear, while the old ones become unstable minima of $\mathcal{L}_{IB}$.}
	\label{fig:fig1}
\end{figure}

Relevant features of any data, physical or not, are only meaningfully defined relative to the task at hand, and their identification is complicated by multiple ``irrelevant" (for the question asked) structures or regularities which may simultaneously exist in the data. 
The Information Bottleneck provides a rigorous framework for \emph{unsupervised learning} of such most relevant features.
With joint probability distribution of ``data" $V$ and an auxiliary ``relevance" variable $E$ as inputs, the IB finds the optimal (lossy) compression $H$ of $V$ preserving information about $E$ (see Fig.\ref{fig:fig1}). The correlations with $E$ thus \emph{define} what is ``relevant" in $V$, rather than arbitrary measures. IB can be posed as the following variational problem:
\begin{equation}\label{eq:main_ib_lag1}
\min_{P(H|V)} \mathcal{L}_{IB}[P(H|V)] \equiv \min_{P(H|V)} I(V;H) - \beta I(H;E),
\end{equation}
where the optimization is over conditional probability distributions $P(H|V)$ describing the encoding of $V$ into $H$. The mutual information terms $I$ in $\mathcal{L}_{IB}$ quantify total retained information (\emph{i.e.}~compression rate), and the relevant information thus preserved, respectively, with parameter $\beta \geq 0$ controlling the tradeoff between them. 

The optimal encoder is found either by writing down a set of coupled IB equations for distributions $P(H|V)$, $P(H)$, $P(E|H)$ and solving them iteratively (see SM, and Ref.\cite{Hassanpour2017} for algorithms), or more practically, applying ML variational inference techniques \cite{AlemiFD016}. For the formal analysis here the IB equations are used; the efficient RSMI-NE algorithm is based on ML methods~\cite{RSMI-NE}. Strikingly, the optimal encoders undergo a sequence of phase transitions as $\beta$ is varied (see Fig.\ref{fig:fig1}). Particularly, the encoder is trivial (retaining zero information) until a \emph{finite} value of $\beta_{c,1}$ at which the first IB transition occurs, when the gain due to retaining some (most) relevant aspect of data outweighs the penalty for keeping any information at all. At each subsequent transition the encoder begins to track another distinct \emph{feature}. This discontinuous behaviour, both for discrete~\cite{parkernips,parkerentropy} and continuous variables~\cite{NIPS2003_2457}, is crucial, allowing to identify such well-defined features.

\begin{figure}[tp]
	\centering
	\includegraphics[width=8.5cm,trim={5cm 4cm 2.5cm 4cm}]{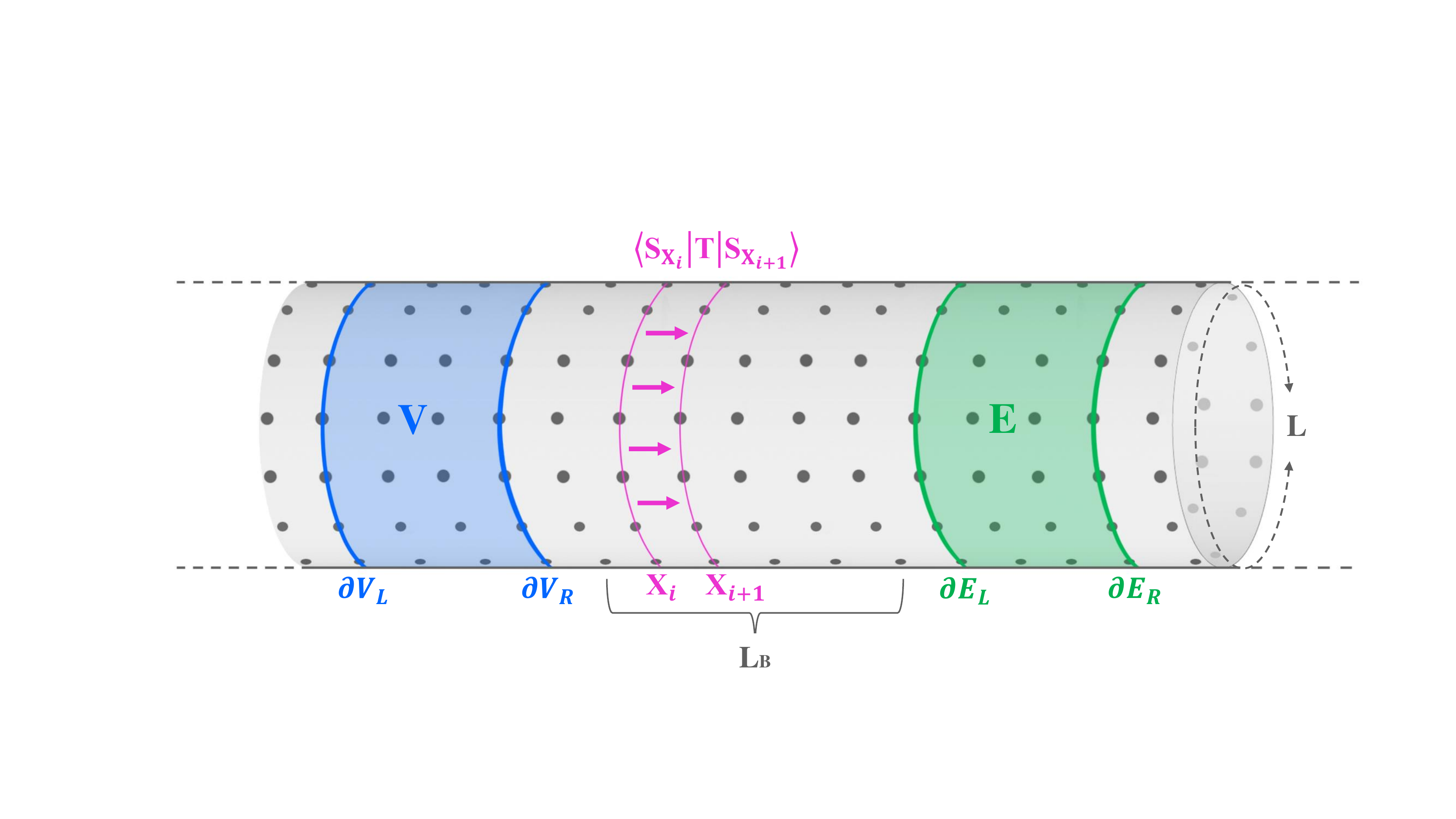}
	\caption{The transfer matrix (TM) setup used. For a system on a cylinder the IB equations can be solved in terms of TM eigenvectors, which are related to the CFT data in the limit of large circumference $L$.}
	\label{fig:transfer2}
\end{figure}

While the IB may be applied to any data, it is of fundamental interest to confront the notion of relevance it gives rise to, and the features it extracts, with \emph{the} physical relevance, as defined by RG. The former being determined by the relevance variable, we need $E$ ensuring the IB retains precisely the RG-relevant information. An appropriate definition for real-space RG was postulated in the context of RSMI \cite{Koch-Janusz2018,PhysRevX.10.011037}: for a random variable $V$  representing the marginal distribution of degrees of freedom in an area to be coarse-grained the variable $E$ (the ``environment"), is the remainder of the system beyond a shell of non-zero thickness around $V$ (the ``buffer", see Fig.\ref{fig:transfer2}). The thickness of the excluded buffer, formally taken to infinity, sets the length scale separating short-range correlations to be discarded, from information about long-range properties of the system. 
Despite conceptual appeal -- the system itself defines relevance -- and partial numerical \cite{Koch-Janusz2018} and theoretical evidence \cite{PhysRevX.10.011037}, the validity of this approach and its relation to RG field-theory formalism were unclear. There are also subtle differences between the IB and RSMI approaches. We now can resolve these issues.

To this end consider a statistical mechanical system on a cylinder; the subsystem to be coarse-grained $V$, the buffer, and the relevance variable $E$ are its subsections as per Fig.\ref{fig:transfer2}. We assume the system is governed by short-range interactions, and use the classic transfer-matrix (TM) method \cite{1941PhRv...60..252K,1944PhRv...65..117O,doi:10.1063/1.330232}: the partition function can be written as $\mathcal{Z} = \langle BC| \mathcal{T}^{L_\infty} | BC \rangle$, where is $L_\infty$ the system length, and the entries of $\mathcal{T}$ are matrix elements of the exponentiated Hamiltonian between configurations of degrees of freedom on elementary slices of the cylinder (in a lattice system; in continuum they are taken between subsequent slices of the states in the discretised path integral picture). We use braket notation for such configurations, in particular $| BC \rangle$ are boundary conditions at the cylinder ends. The unique advantage of the TM approach is that, on the one hand, all distributions entering the IB equations can be cast as matrix elements and partial traces of powers of $\mathcal{T}$, and on the other hand the eigenvalues $\lambda_i$ and eigenvectors $|i\rangle$ of $\mathcal{T}$ have a direct relation to the operator content of the CFT describing the system \cite{latticeanimals,Cardy_1984,CARDY1986186}. Specifically, $\lambda_i/\lambda_0 = e^{-\frac{2\pi}{L}\Delta_i}$ in the limit of large cylinder circumference $L$, where $\Delta_i$ are the total scaling dimensions of the CFT primaries (determining the RG scaling dimensions, and so the critical exponents) in ascending order. TM thus serves as a theoretical dictionary helping to establish a \emph{quantitative} map between the field- and information-theoretic objects. 

\captionsetup[figure]{justification=raggedright}
To be concrete, consider the IB equations for the optimal encoder $P(h|v)$ at fixed tradeoff $\beta$ (see SM and \cite{infbottle1}):
\begin{align}\label{eq:ibeqs1_maintext}
P(h|v) &\propto P(h) e^{\beta \sum_{e} P(e|v)\log( P(e|h))} \\ \nonumber 
P(e|h) &= \sum\nolimits_{v} p(e|v)p(v|h),
\end{align}
where $e$, $h$, $v$ are configurations of ${\rm E}$, ${\rm H}$ and ${\rm V}$. Observe first that the distribution $P(X)$ of any cylindrical section $X$ of the system (\emph{e.g.}~V or E) can be written using $\mathcal{T}$:
\begin{align} \nonumber
P(X) = \langle 0| \partial x_L \rangle\langle \partial x_L| \mathcal{T}|x_2\rangle\langle x_2 | \mathcal{T}\ldots \mathcal{T}|\partial x_R\rangle \langle \partial x_R| 0\rangle
\end{align}
Here $x_i$ are successive slices of $X$; the configurations of the boundary slices  are denoted as $\partial x_{R/L}$ and $|0\rangle = \mathcal{T}^{L_{\infty}}|BC\rangle $ is the CFT vacuum, on which $\mathcal{T}$ acts as an identity. We used the eigendecomposition $\mathcal{T} = |0\rangle\langle 0| + \sum_{i} e^{-2\pi\Delta_i /L} |\Delta_i\rangle\langle \Delta_i|$, with $|\Delta_i\rangle = \phi_{\Delta_i}|0\rangle$ created by primary fields $\phi_{\Delta_i}$ with conformal dimension $\Delta_i$. All distributions in Eqs.\ref{eq:ibeqs1_maintext} can be written analogously as functions of $\mathcal{T}$, using the eigendecomposition and Bayes' law, phrasing the IB equations fully in TM terms.

Eqs.\ref{eq:ibeqs1_maintext} are highly non-linear and coupled. Remarkably though, the only interdependence of the conditional probabilities in the large buffer limit is, as shown in SM, via:
\begin{align}\label{eq:rxre1_maintext}
r_v = \frac{\langle 0| \phi_{\Delta_1} |\partial V_R \rangle}{\langle 0 | \partial V_R \rangle} \mbox{\ \ \ \ \ \ }r_e = \frac{\langle \partial E_L|\phi_{\Delta_1} |0 \rangle}{\langle \partial E_L | 0 \rangle},
\end{align}
\emph{i.e.}~the matrix elements of the CFT primary fields of lowest scaling dimensions. In particular, we prove that:
\begin{align}
\label{Eq:reducedIB_maintext}
P(h|v) = P(h|r_v) &\propto P(h) e^{\beta \epsilon^2 r_v \langle r_v \rangle_{h}}, 
\end{align}
with $\epsilon = (\lambda_1/\lambda_0)^{L_B}$.

Eq.\ref{Eq:reducedIB_maintext} is one of our key results: the optimal IB encoder depends on $V$ \emph{only} via $r_v$, \emph{i.e.}~the matrix element of the most relevant operator in the sense of RG (the dependence on the variance of $r_e$ can be absorbed into rescaling of $\beta$, and in $\langle r_v \rangle_{h}$ v-dependence is averaged over). The solution changes from one system (CFT) to another through the values of $r_v$ and $\langle r_v \rangle_{h}$. 
This is the mathematical statement of the equivalence of the IB and RG relevance. In other words, the ``features" the IB, and consequently the RSMI, extract are not arbitrary, but correspond to physically most relevant operators.

Though Eq.\ref{Eq:reducedIB_maintext} is implicit, as $\langle r_v \rangle_{h}$ depends on $P(h|v)$, it can be analytically solved around the first IB transition, \emph{i.e.}~for $\beta = \beta_{c,1} + t$. Below $\beta_{c,1}$ no information is retained: the encoder is independent of $V$ and trivial: $P(h|r_v) = 1/|H|$, with $|H|$ the cardinality of the coarse-grained variable. Equiprobability of $h$ reflects a structural symmetry of the encoder under permutations of $h$ labels. Any nontrivial encoder \emph{must} break it, introducing dependence of some $h$ on $V$ to preserve information: $\beta_{c,1}$ marks the first such breaking (in fact all IB transitions reflect successive breaking of permutation symmetry). 
Above $\beta_{c,1}$, following Refs.\cite{parkernips,parkerentropy}, the encoder can be perturbatively expanded around the trivial solution (see SM for detailed discussion). In particular, comparing to the expansion of Eq.\ref{Eq:reducedIB_maintext} in $t$ yields:
\begin{align}\label{eq:betac1_maintext}
\beta^{-1}_{c,1} &= \epsilon^2 + o(\epsilon^2) \xrightarrow[]{L \rightarrow \infty} e^{-4\pi \Delta_1 \frac{L_B}{L}} + o(\epsilon^2).
\end{align}
Here $o(\epsilon^2)$, containing powers of $\epsilon$ greater than two, reflects the contribution of operators of subleading relevance. As $\epsilon$ decays exponentially in $L_B/L$, maintaining $L_B \gg L$ suppresses these corrections exponentially. 

\begin{figure}[t]
	
	\includegraphics[width=\columnwidth,trim={0 0 0 1.5cm}]{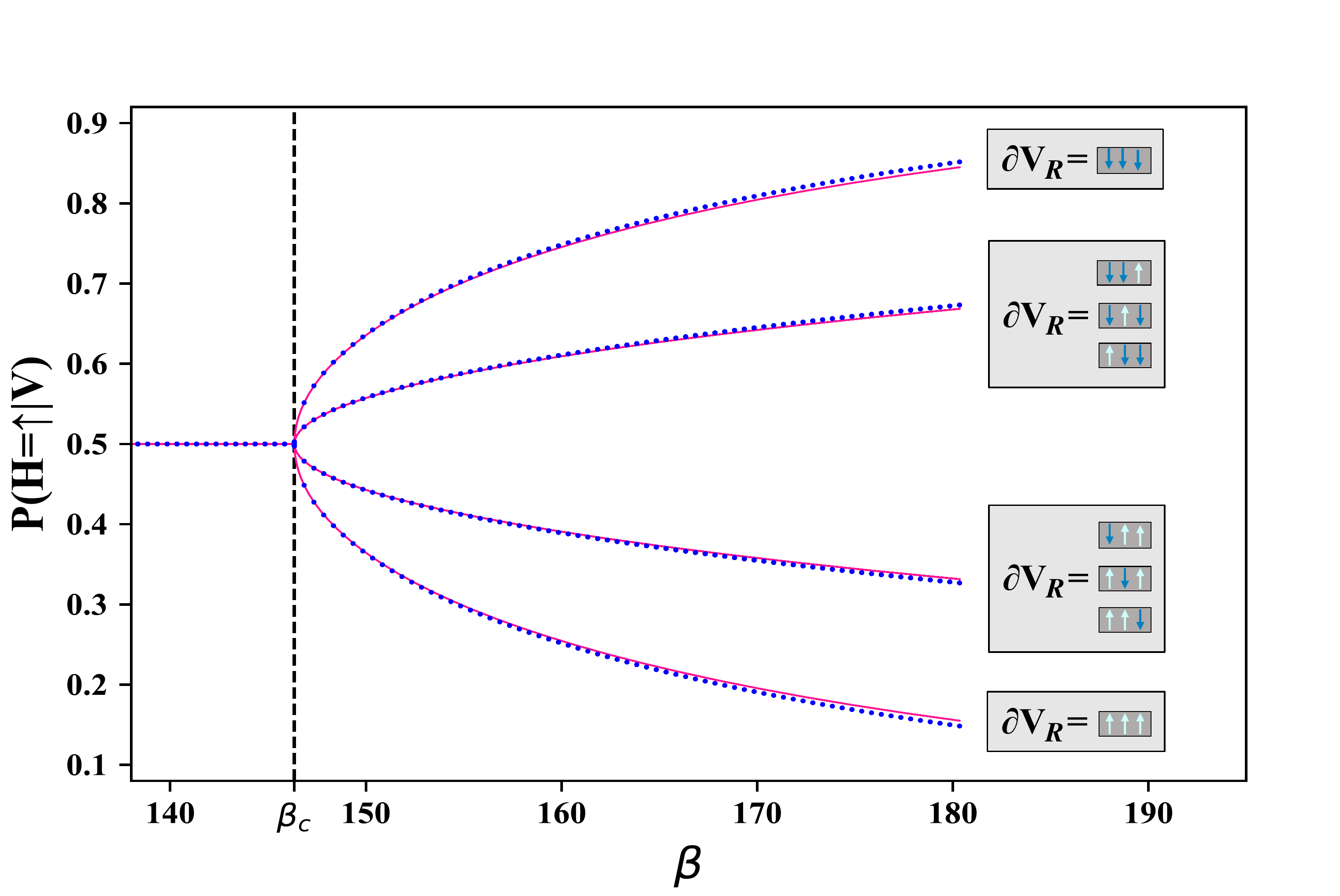}
\caption{Comparison of theory with numerics. For the critical 2D Ising system, the analytical prediction (solid red) for the optimal compression $P_\beta(h|v)$ (see Eq.\ref{Eq:reducedIB_maintext}, and Eq.\ref{eq:encoder1} in SM) is confronted with encoders obtained by numerically solving IB Eqs.\ref{eq:ibeqs1_maintext} on the probability distribution of the system (blue dots). For clarity we use a cylinder of three sites' circumference, $V$ and $E$ as in Fig.\ref{fig:transfer2}. The variable $H$ is a spin, whose probability to take value $\uparrow$ we plot as a function of the tradeoff $\beta$ (see Eq.\ref{eq:main_ib_lag1}). The encoder is completely random and independent of $V$ below $\beta_c$ matching the prediction Eq.\ref{eq:betac1_maintext}, and above is determined by the magnetization on the edge, in excellent agreement with the theory.}
	\label{fig:fig3}
\end{figure}

Equation \ref{eq:betac1_maintext} is an analytical prediction for the IB phase transition, signaling emergence of nontrivial solutions to the IB equations (see Fig.\ref{fig:fig1} and Fig.\ref{fig:ibrsmi} in SM), in terms of field-theoretic quantities characterizing the physical system. In SM, utilizing the structure of the Hessian of $\mathcal{L}_{IB}$, we also derive this solution explicitly (see also Fig.\ref{fig:fig3}).

The prediction is generic and verifiable: we can input the probability distribution of the system to the IB equations, and find the solutions for changing $\beta$ numerically, as in a compression problem \cite{Hassanpour2017}. On the other hand we can use the CFT description 
and either compute $r_v$,  $\langle r_v \rangle_{h}$ and $\epsilon$ analytically, or by a numerical transfer matrix diagonalization, and compare. In Fig.\ref{fig:fig3}c numerical IB solutions are plotted as a function of $\beta$ in the case of critical 2D Ising model. The value $\beta_{c,1}^{IB}$ at which non-trivial encoders appear matches the predicted $\beta_{c,1}$ to high accuracy. The feature the IB extracts is indeed the most relevant local operator, \emph{i.e.}~the magnetization (see SM).

The validity of this picture is not limited to lattice models. In fact, for the continuum Gaussian field theory the entire IB curve can be computed analytically, including \emph{all} the IB phase transitions \cite{aditya}, using Gaussian Information Bottleneck results~\cite{NIPS2003_2457} and Green's functions.

\captionsetup[figure]{justification=raggedright}
As mentioned, the RSMI algorithm \cite{Koch-Janusz2018,PhysRevX.10.011037} is closely related to the IB. Specifically, it also maximizes the relevant information $I(H;E)$, however contains no tradeoff $\beta$, but instead a fixed cardinality $|H|$. Intuitively, the IB extracts as meany features as $\beta$ allows, adding them as $\beta$ grows, while the RSMI from the outset optimizes exactly $|H|$ best features. RSMI is thus a $\beta \rightarrow \infty$ limit of IB under the constraint of fixed $|H|$. 
In practice $|H|$ is also bounded in IB, however this affects solutions only at $\beta$ large enough for $|H|$ features to have already been used.

The quantitative connection between compression- and field-theoretic formalisms thus established opens the exciting possibility of applying distinct theoretical and numerical methods of either area to its counterpart. We discuss such avenues in the conclusions, here, however, we immediately demonstrate one interesting example.

Symmetries are crucial in analytical understanding of physical systems, and in RG in particular \cite{zinn1989quantum}. They have a direct relation to order parameters, and often effectively determine the long range properties. One thus expects IB and RSMI to reflect the relevant symmetries of the model. 
Let $s$ be an element of such symmetry group $\mathcal{S}$ acting on configurations of $V$ and $E$ as a permutation, denoted by multiplication, leaving the system invariant: $P(e,v) = P(se,sv)$. 
We expect the optimal encoder $P_\beta(h|v)$ to maintain it:
\begin{align}\label{Eq:sym_rel}
P(e,v) = P(se,sv) &\Rightarrow P_\beta(h|v) = P_\beta(\phi_s h | s v),
\end{align}
so that the coarse-grained system is invariant under a representation $\phi_s$ of $\mathcal{S}$, potentially trivial. 
We show this indeed holds true in IB, as long as $|H|$ is large enough to support a representation of an appropriate dimension. The argument, detailed in SM, is constructive: below $\beta_{c,1}$ the encoder is trivially invariant under all symmetries. For $\beta = \beta_{c,1} + t$ a solution can be built by an explicit symmetrization procedure, utilizing the knowledge of the perturbative structure of the encoder and the Hessian of $\mathcal{L}_{IB}$ around the first IB transition \cite{parkerentropy}. We show this solution to be optimal. The symmetry of the encoder will hold for all $\beta < \beta_{c,2}$ by continuity; numerical experiments support validity of this picture also more generally.

Note that the symmetry $\mathcal{S}$ may not be obvious in the microscopic formulation of the system \cite{Zhang1990} or the experimental data, or may even be emergent \cite{Senthil2004}. Eq.~\ref{Eq:sym_rel} can then be used as a constructive tool, potentially allowing to systematically learn $\mathcal{S}$ from the symmetries of the entries of the numerically obtained $P_\beta(h|v)$ (SM, see also \cite{Bondesan2019}). Moreover, the structure of the IB in the presence of physical/data symmetries shines light on the question of constructing RG transformations compatible with the symmetries of the system.

The results we presented, though requiring some level of technicality, have clear theoretical interpretations. In fact, their very point is to formalize concepts and connections which ought to be intuitive, and to give the necessary technology to make those quantitative and computable analytically and numerically. Consequently, numerous directions are now open. On a theoretical front, application of IB analysis to extract relevant quantities in the challenging case of disordered and non-equilibrium systems is extremely promising, given its non-reliance on the notion of a Hamiltonian. This may require deeper understanding of the properties of the IB equations, and their constrained version in the RSMI-NE algorithm. Numerically, given the relation to the transfer matrix, the possibility of using the IB/RSMI where TM computations are difficult (\emph{e.g.}~in 3D) is an exciting prospect, as is applying approximate numerical IB or RSMI to experimental data. Finally, we hope that the methodology of using information-theoretical formulations of physical quantities combined with the ability of deep learning to optimize them in a controlled fashion \cite{RSMI-NE}, can provide a blueprint for more theoretically interpretable applications of deep learning in physics.

\textbf{Acknowledgements:} ZR and AG acknowledge support from ISF grant 2250/19 as well as helpful discussions with Etam Benger (HUJI). MKJ gratefully acknowledges the support of Sebastian Huber, during stay in whose group at ETH Zurich part of the work was performed, the financial support from the Swiss National Science Foundation and the NCCR QSIT, and from the European Research Council under the Grant Agreement No.~771503 (TopMechMat), as well as from European Union’s Horizon 2020 programme under Marie Sklodowska-Curie Grant Agreement No.~896004 (COMPLEX ML). AB acknowledges financial support from Eldad Bettelheim's ISF grant 1466/15 and Alex Retzker's MicroQC grant number 820314.

\newpage
\bibliography{ib_bib2}{}

\begin{thebibliography}{59}%
\makeatletter
\providecommand \@ifxundefined [1]{%
 \@ifx{#1\undefined}
}%
\providecommand \@ifnum [1]{%
 \ifnum #1\expandafter \@firstoftwo
 \else \expandafter \@secondoftwo
 \fi
}%
\providecommand \@ifx [1]{%
 \ifx #1\expandafter \@firstoftwo
 \else \expandafter \@secondoftwo
 \fi
}%
\providecommand \natexlab [1]{#1}%
\providecommand \enquote  [1]{``#1''}%
\providecommand \bibnamefont  [1]{#1}%
\providecommand \bibfnamefont [1]{#1}%
\providecommand \citenamefont [1]{#1}%
\providecommand \href@noop [0]{\@secondoftwo}%
\providecommand \href [0]{\begingroup \@sanitize@url \@href}%
\providecommand \@href[1]{\@@startlink{#1}\@@href}%
\providecommand \@@href[1]{\endgroup#1\@@endlink}%
\providecommand \@sanitize@url [0]{\catcode `\\12\catcode `\$12\catcode
  `\&12\catcode `\#12\catcode `\^12\catcode `\_12\catcode `\%12\relax}%
\providecommand \@@startlink[1]{}%
\providecommand \@@endlink[0]{}%
\providecommand \url  [0]{\begingroup\@sanitize@url \@url }%
\providecommand \@url [1]{\endgroup\@href {#1}{\urlprefix }}%
\providecommand \urlprefix  [0]{URL }%
\providecommand \Eprint [0]{\href }%
\providecommand \doibase [0]{http://dx.doi.org/}%
\providecommand \selectlanguage [0]{\@gobble}%
\providecommand \bibinfo  [0]{\@secondoftwo}%
\providecommand \bibfield  [0]{\@secondoftwo}%
\providecommand \translation [1]{[#1]}%
\providecommand \BibitemOpen [0]{}%
\providecommand \bibitemStop [0]{}%
\providecommand \bibitemNoStop [0]{.\EOS\space}%
\providecommand \EOS [0]{\spacefactor3000\relax}%
\providecommand \BibitemShut  [1]{\csname bibitem#1\endcsname}%
\let\auto@bib@innerbib\@empty
\bibitem [{\citenamefont {Coveney}\ \emph {et~al.}(2016)\citenamefont
  {Coveney}, \citenamefont {Dougherty},\ and\ \citenamefont
  {Highfield}}]{doi:10.1098/rsta.2016.0153}%
  \BibitemOpen
  \bibfield  {author} {\bibinfo {author} {\bibfnamefont {Peter~V.}\
  \bibnamefont {Coveney}}, \bibinfo {author} {\bibfnamefont {Edward~R.}\
  \bibnamefont {Dougherty}}, \ and\ \bibinfo {author} {\bibfnamefont
  {Roger~R.}\ \bibnamefont {Highfield}},\ }\bibfield  {title} {\enquote
  {\bibinfo {title} {Big data need big theory too},}\ }\href {\doibase
  10.1098/rsta.2016.0153} {\bibfield  {journal} {\bibinfo  {journal}
  {Philosophical Transactions of the Royal Society A: Mathematical, Physical
  and Engineering Sciences}\ }\textbf {\bibinfo {volume} {374}} (\bibinfo
  {year} {2016}),\ 10.1098/rsta.2016.0153}\BibitemShut {NoStop}%
\bibitem [{\citenamefont {Carleo}\ \emph {et~al.}(2019)\citenamefont {Carleo},
  \citenamefont {Cirac}, \citenamefont {Cranmer}, \citenamefont {Daudet},
  \citenamefont {Schuld}, \citenamefont {Tishby}, \citenamefont
  {Vogt-Maranto},\ and\ \citenamefont {Zdeborov\'a}}]{RevModPhys.91.045002}%
  \BibitemOpen
  \bibfield  {author} {\bibinfo {author} {\bibfnamefont {Giuseppe}\
  \bibnamefont {Carleo}}, \bibinfo {author} {\bibfnamefont {Ignacio}\
  \bibnamefont {Cirac}}, \bibinfo {author} {\bibfnamefont {Kyle}\ \bibnamefont
  {Cranmer}}, \bibinfo {author} {\bibfnamefont {Laurent}\ \bibnamefont
  {Daudet}}, \bibinfo {author} {\bibfnamefont {Maria}\ \bibnamefont {Schuld}},
  \bibinfo {author} {\bibfnamefont {Naftali}\ \bibnamefont {Tishby}}, \bibinfo
  {author} {\bibfnamefont {Leslie}\ \bibnamefont {Vogt-Maranto}}, \ and\
  \bibinfo {author} {\bibfnamefont {Lenka}\ \bibnamefont {Zdeborov\'a}},\
  }\bibfield  {title} {\enquote {\bibinfo {title} {Machine learning and the
  physical sciences},}\ }\href {\doibase 10.1103/RevModPhys.91.045002}
  {\bibfield  {journal} {\bibinfo  {journal} {Rev. Mod. Phys.}\ }\textbf
  {\bibinfo {volume} {91}},\ \bibinfo {pages} {045002} (\bibinfo {year}
  {2019})}\BibitemShut {NoStop}%
\bibitem [{\citenamefont {Murdoch}\ \emph {et~al.}(2019)\citenamefont
  {Murdoch}, \citenamefont {Singh}, \citenamefont {Kumbier}, \citenamefont
  {Abbasi-Asl},\ and\ \citenamefont {Yu}}]{Murdoch22071}%
  \BibitemOpen
  \bibfield  {author} {\bibinfo {author} {\bibfnamefont {W.~James}\
  \bibnamefont {Murdoch}}, \bibinfo {author} {\bibfnamefont {Chandan}\
  \bibnamefont {Singh}}, \bibinfo {author} {\bibfnamefont {Karl}\ \bibnamefont
  {Kumbier}}, \bibinfo {author} {\bibfnamefont {Reza}\ \bibnamefont
  {Abbasi-Asl}}, \ and\ \bibinfo {author} {\bibfnamefont {Bin}\ \bibnamefont
  {Yu}},\ }\bibfield  {title} {\enquote {\bibinfo {title} {Definitions,
  methods, and applications in interpretable machine learning},}\ }\href
  {\doibase 10.1073/pnas.1900654116} {\bibfield  {journal} {\bibinfo  {journal}
  {Proceedings of the National Academy of Sciences}\ }\textbf {\bibinfo
  {volume} {116}},\ \bibinfo {pages} {22071--22080} (\bibinfo {year}
  {2019})}\BibitemShut {NoStop}%
\bibitem [{\citenamefont {Preskill}(2000)}]{preskill}%
  \BibitemOpen
  \bibfield  {author} {\bibinfo {author} {\bibfnamefont {John}\ \bibnamefont
  {Preskill}},\ }\bibfield  {title} {\enquote {\bibinfo {title} {Quantum
  information and physics: Some future directions},}\ }\href {\doibase
  10.1080/09500340008244031} {\bibfield  {journal} {\bibinfo  {journal}
  {Journal of Modern Optics}\ }\textbf {\bibinfo {volume} {47}},\ \bibinfo
  {pages} {127--137} (\bibinfo {year} {2000})}\BibitemShut {NoStop}%
\bibitem [{\citenamefont {MacKay}(2002)}]{10.5555/971143}%
  \BibitemOpen
  \bibfield  {author} {\bibinfo {author} {\bibfnamefont {David J.~C.}\
  \bibnamefont {MacKay}},\ }\href@noop {} {\emph {\bibinfo {title}
  {{Information Theory, Inference, and Learning Algorithms}}}}\ (\bibinfo
  {publisher} {Cambridge University Press},\ \bibinfo {year}
  {2002})\BibitemShut {NoStop}%
\bibitem [{\citenamefont {Mezard}\ and\ \citenamefont
  {Montanari}(2009)}]{10.5555/1592967}%
  \BibitemOpen
  \bibfield  {author} {\bibinfo {author} {\bibfnamefont {Marc}\ \bibnamefont
  {Mezard}}\ and\ \bibinfo {author} {\bibfnamefont {Andrea}\ \bibnamefont
  {Montanari}},\ }\href@noop {} {\emph {\bibinfo {title} {{Information,
  Physics, and Computation}}}}\ (\bibinfo  {publisher} {Oxford University
  Press, Inc.},\ \bibinfo {address} {USA},\ \bibinfo {year} {2009})\BibitemShut
  {NoStop}%
\bibitem [{\citenamefont {Shannon}(1948)}]{Shannon48}%
  \BibitemOpen
  \bibfield  {author} {\bibinfo {author} {\bibfnamefont {Claude~E.}\
  \bibnamefont {Shannon}},\ }\bibfield  {title} {\enquote {\bibinfo {title} {A
  mathematical theory of communication.}}\ }\href@noop {} {\bibfield  {journal}
  {\bibinfo  {journal} {Bell Syst. Tech. J.}\ }\textbf {\bibinfo {volume}
  {27}},\ \bibinfo {pages} {379--423} (\bibinfo {year} {1948})}\BibitemShut
  {NoStop}%
\bibitem [{\citenamefont {Blau}\ and\ \citenamefont
  {Michaeli}(2019)}]{BlauM19}%
  \BibitemOpen
  \bibfield  {author} {\bibinfo {author} {\bibfnamefont {Yochai}\ \bibnamefont
  {Blau}}\ and\ \bibinfo {author} {\bibfnamefont {Tomer}\ \bibnamefont
  {Michaeli}},\ }\bibfield  {title} {\enquote {\bibinfo {title} {{Rethinking
  Lossy Compression: The Rate-Distortion-Perception Tradeoff}},}\ }in\
  \href@noop {} {\emph {\bibinfo {booktitle} {Proceedings of the 36th
  International Conference on Machine Learning, {ICML} 2019}}},\ \bibinfo
  {series} {Proceedings of Machine Learning Research}, Vol.~\bibinfo {volume}
  {97}\ (\bibinfo  {publisher} {{PMLR}},\ \bibinfo {year} {2019})\ pp.\
  \bibinfo {pages} {675--685}\BibitemShut {NoStop}%
\bibitem [{\citenamefont {{Tishby}}\ \emph {et~al.}(2001)\citenamefont
  {{Tishby}}, \citenamefont {{Pereira}},\ and\ \citenamefont
  {{Bialek}}}]{infbottle1}%
  \BibitemOpen
  \bibfield  {author} {\bibinfo {author} {\bibfnamefont {N.}~\bibnamefont
  {{Tishby}}}, \bibinfo {author} {\bibfnamefont {F.~C.}\ \bibnamefont
  {{Pereira}}}, \ and\ \bibinfo {author} {\bibfnamefont {W.}~\bibnamefont
  {{Bialek}}},\ }\bibfield  {title} {\enquote {\bibinfo {title} {{The
  information bottleneck method}},}\ }\bibfield  {booktitle} {\emph {\bibinfo
  {booktitle} {Proceedings of the 37th Allerton Conference on Communication,
  Control and Computation}},\ }\href@noop {} {\ \textbf {\bibinfo {volume}
  {49}} (\bibinfo {year} {2001})}\BibitemShut {NoStop}%
\bibitem [{\citenamefont {Alemi}\ \emph {et~al.}(2016)\citenamefont {Alemi},
  \citenamefont {Fischer}, \citenamefont {Dillon},\ and\ \citenamefont
  {Murphy}}]{AlemiFD016}%
  \BibitemOpen
  \bibfield  {author} {\bibinfo {author} {\bibfnamefont {Alexander~A.}\
  \bibnamefont {Alemi}}, \bibinfo {author} {\bibfnamefont {Ian}\ \bibnamefont
  {Fischer}}, \bibinfo {author} {\bibfnamefont {Joshua~V.}\ \bibnamefont
  {Dillon}}, \ and\ \bibinfo {author} {\bibfnamefont {Kevin}\ \bibnamefont
  {Murphy}},\ }\bibfield  {title} {\enquote {\bibinfo {title} {Deep variational
  information bottleneck},}\ }\href {http://arxiv.org/abs/1612.00410}
  {\bibfield  {journal} {\bibinfo  {journal} {CoRR}\ }\textbf {\bibinfo
  {volume} {abs/1612.00410}} (\bibinfo {year} {2016})},\ \Eprint
  {http://arxiv.org/abs/1612.00410} {arXiv:1612.00410} \BibitemShut {NoStop}%
\bibitem [{\citenamefont {Wilson}\ and\ \citenamefont
  {Kogut}(1974)}]{Wilson1974}%
  \BibitemOpen
  \bibfield  {author} {\bibinfo {author} {\bibfnamefont {Kenneth~G.}\
  \bibnamefont {Wilson}}\ and\ \bibinfo {author} {\bibfnamefont {John}\
  \bibnamefont {Kogut}},\ }\bibfield  {title} {\enquote {\bibinfo {title} {The
  renormalization group and the $\epsilon$ expansion},}\ }\href {\doibase
  https://doi.org/10.1016/0370-1573(74)90023-4} {\bibfield  {journal} {\bibinfo
   {journal} {Physics Reports}\ }\textbf {\bibinfo {volume} {12}},\ \bibinfo
  {pages} {75 -- 199} (\bibinfo {year} {1974})}\BibitemShut {NoStop}%
\bibitem [{\citenamefont {Wilson}(1975)}]{Wilson1975}%
  \BibitemOpen
  \bibfield  {author} {\bibinfo {author} {\bibfnamefont {Kenneth~G.}\
  \bibnamefont {Wilson}},\ }\bibfield  {title} {\enquote {\bibinfo {title}
  {{The renormalization group: Critical phenomena and the Kondo problem}},}\
  }\href {\doibase 10.1103/RevModPhys.47.773} {\bibfield  {journal} {\bibinfo
  {journal} {Rev. Mod. Phys.}\ }\textbf {\bibinfo {volume} {47}},\ \bibinfo
  {pages} {773--840} (\bibinfo {year} {1975})}\BibitemShut {NoStop}%
\bibitem [{\citenamefont {Fisher}(1998)}]{Fisher1998}%
  \BibitemOpen
  \bibfield  {author} {\bibinfo {author} {\bibfnamefont {Michael~E.}\
  \bibnamefont {Fisher}},\ }\bibfield  {title} {\enquote {\bibinfo {title}
  {Renormalization group theory: Its basis and formulation in statistical
  physics},}\ }\href {\doibase 10.1103/RevModPhys.70.653} {\bibfield  {journal}
  {\bibinfo  {journal} {Rev. Mod. Phys.}\ }\textbf {\bibinfo {volume} {70}},\
  \bibinfo {pages} {653--681} (\bibinfo {year} {1998})}\BibitemShut {NoStop}%
\bibitem [{\citenamefont {Belavin}\ \emph
  {et~al.}(1984{\natexlab{a}})\citenamefont {Belavin}, \citenamefont
  {Polyakov},\ and\ \citenamefont {Zamolodchikov}}]{BELAVIN_first}%
  \BibitemOpen
  \bibfield  {author} {\bibinfo {author} {\bibfnamefont {A.A.}\ \bibnamefont
  {Belavin}}, \bibinfo {author} {\bibfnamefont {A.M.}\ \bibnamefont
  {Polyakov}}, \ and\ \bibinfo {author} {\bibfnamefont {A.B.}\ \bibnamefont
  {Zamolodchikov}},\ }\bibfield  {title} {\enquote {\bibinfo {title} {{Infinite
  conformal symmetry of critical fluctuations in two dimensions}},}\
  }\href@noop {} {\bibfield  {journal} {\bibinfo  {journal} {Journal of
  Statistical Physics}\ }\textbf {\bibinfo {volume} {34}},\ \bibinfo {pages}
  {763--774} (\bibinfo {year} {1984}{\natexlab{a}})}\BibitemShut {NoStop}%
\bibitem [{\citenamefont {Belavin}\ \emph
  {et~al.}(1984{\natexlab{b}})\citenamefont {Belavin}, \citenamefont
  {Polyakov},\ and\ \citenamefont {Zamolodchikov}}]{BELAVIN1984333}%
  \BibitemOpen
  \bibfield  {author} {\bibinfo {author} {\bibfnamefont {A.A.}\ \bibnamefont
  {Belavin}}, \bibinfo {author} {\bibfnamefont {A.M.}\ \bibnamefont
  {Polyakov}}, \ and\ \bibinfo {author} {\bibfnamefont {A.B.}\ \bibnamefont
  {Zamolodchikov}},\ }\bibfield  {title} {\enquote {\bibinfo {title} {{Infinite
  conformal symmetry in two-dimensional quantum field theory}},}\ }\href
  {\doibase https://doi.org/10.1016/0550-3213(84)90052-X} {\bibfield  {journal}
  {\bibinfo  {journal} {Nuclear Physics B}\ }\textbf {\bibinfo {volume}
  {241}},\ \bibinfo {pages} {333 -- 380} (\bibinfo {year}
  {1984}{\natexlab{b}})}\BibitemShut {NoStop}%
\bibitem [{\citenamefont {Friedan}\ \emph {et~al.}(1984)\citenamefont
  {Friedan}, \citenamefont {Qiu},\ and\ \citenamefont
  {Shenker}}]{PhysRevLett.52.1575}%
  \BibitemOpen
  \bibfield  {author} {\bibinfo {author} {\bibfnamefont {Daniel}\ \bibnamefont
  {Friedan}}, \bibinfo {author} {\bibfnamefont {Zongan}\ \bibnamefont {Qiu}}, \
  and\ \bibinfo {author} {\bibfnamefont {Stephen}\ \bibnamefont {Shenker}},\
  }\bibfield  {title} {\enquote {\bibinfo {title} {{Conformal Invariance,
  Unitarity, and Critical Exponents in Two Dimensions}},}\ }\href {\doibase
  10.1103/PhysRevLett.52.1575} {\bibfield  {journal} {\bibinfo  {journal}
  {Phys. Rev. Lett.}\ }\textbf {\bibinfo {volume} {52}},\ \bibinfo {pages}
  {1575--1578} (\bibinfo {year} {1984})}\BibitemShut {NoStop}%
\bibitem [{\citenamefont {Di~Francesco}\ \emph {et~al.}(1997)\citenamefont
  {Di~Francesco}, \citenamefont {Mathieu},\ and\ \citenamefont
  {Sénéchal}}]{DiFrancesco:639405}%
  \BibitemOpen
  \bibfield  {author} {\bibinfo {author} {\bibfnamefont {Philippe}\
  \bibnamefont {Di~Francesco}}, \bibinfo {author} {\bibfnamefont {Pierre}\
  \bibnamefont {Mathieu}}, \ and\ \bibinfo {author} {\bibfnamefont {David}\
  \bibnamefont {Sénéchal}},\ }\href {\doibase 10.1007/978-1-4612-2256-9}
  {\emph {\bibinfo {title} {{Conformal field theory}}}},\ Graduate texts in
  contemporary physics\ (\bibinfo  {publisher} {Springer},\ \bibinfo {address}
  {New York, NY},\ \bibinfo {year} {1997})\BibitemShut {NoStop}%
\bibitem [{\citenamefont {Cardy}(1996)}]{Cardy:318508}%
  \BibitemOpen
  \bibfield  {author} {\bibinfo {author} {\bibfnamefont {John~L}\ \bibnamefont
  {Cardy}},\ }\href@noop {} {\emph {\bibinfo {title} {{Scaling and
  renormalization in statistical physics}}}},\ Cambridge lectrue notes in
  physics\ (\bibinfo  {publisher} {Cambridge Univ. Press},\ \bibinfo {address}
  {Cambridge},\ \bibinfo {year} {1996})\BibitemShut {NoStop}%
\bibitem [{\citenamefont {Itzykson}\ \emph {et~al.}(1998)\citenamefont
  {Itzykson}, \citenamefont {Saleur},\ and\ \citenamefont
  {Zuber}}]{doi:10.1142/0608}%
  \BibitemOpen
  \bibfield  {author} {\bibinfo {author} {\bibfnamefont {C.}~\bibnamefont
  {Itzykson}}, \bibinfo {author} {\bibfnamefont {H.}~\bibnamefont {Saleur}}, \
  and\ \bibinfo {author} {\bibfnamefont {J.-B.}\ \bibnamefont {Zuber}},\ }\href
  {\doibase 10.1142/0608} {\emph {\bibinfo {title} {{Conformal Invariance and
  Applications to Statistical Mechanics}}}}\ (\bibinfo  {publisher} {World
  Scientific},\ \bibinfo {year} {1998})\BibitemShut {NoStop}%
\bibitem [{\citenamefont {Poland}\ \emph {et~al.}(2019)\citenamefont {Poland},
  \citenamefont {Rychkov},\ and\ \citenamefont {Vichi}}]{RevModPhys.91.015002}%
  \BibitemOpen
  \bibfield  {author} {\bibinfo {author} {\bibfnamefont {David}\ \bibnamefont
  {Poland}}, \bibinfo {author} {\bibfnamefont {Slava}\ \bibnamefont {Rychkov}},
  \ and\ \bibinfo {author} {\bibfnamefont {Alessandro}\ \bibnamefont {Vichi}},\
  }\bibfield  {title} {\enquote {\bibinfo {title} {{The conformal bootstrap:
  Theory, numerical techniques, and applications}},}\ }\href {\doibase
  10.1103/RevModPhys.91.015002} {\bibfield  {journal} {\bibinfo  {journal}
  {Rev. Mod. Phys.}\ }\textbf {\bibinfo {volume} {91}},\ \bibinfo {pages}
  {015002} (\bibinfo {year} {2019})}\BibitemShut {NoStop}%
\bibitem [{\citenamefont {Zamolodchikov}(1986)}]{Zamolodchikov:1986gt}%
  \BibitemOpen
  \bibfield  {author} {\bibinfo {author} {\bibfnamefont {A.B.}\ \bibnamefont
  {Zamolodchikov}},\ }\bibfield  {title} {\enquote {\bibinfo {title}
  {{Irreversibility of the Flux of the Renormalization Group in a 2D Field
  Theory}},}\ }\href@noop {} {\bibfield  {journal} {\bibinfo  {journal} {JETP
  Lett.}\ }\textbf {\bibinfo {volume} {43}},\ \bibinfo {pages} {730--732}
  (\bibinfo {year} {1986})}\BibitemShut {NoStop}%
\bibitem [{\citenamefont {Gaite}\ and\ \citenamefont
  {O'Connor}(1996)}]{gaite-oconnor}%
  \BibitemOpen
  \bibfield  {author} {\bibinfo {author} {\bibfnamefont {Jos\'e}\ \bibnamefont
  {Gaite}}\ and\ \bibinfo {author} {\bibfnamefont {Denjoe}\ \bibnamefont
  {O'Connor}},\ }\bibfield  {title} {\enquote {\bibinfo {title} {Field theory
  entropy, the $h$ theorem, and the renormalization group},}\ }\href {\doibase
  10.1103/PhysRevD.54.5163} {\bibfield  {journal} {\bibinfo  {journal} {Phys.
  Rev. D}\ }\textbf {\bibinfo {volume} {54}},\ \bibinfo {pages} {5163--5173}
  (\bibinfo {year} {1996})}\BibitemShut {NoStop}%
\bibitem [{\citenamefont {Casini}\ and\ \citenamefont
  {Huerta}(2007)}]{Casini_2007}%
  \BibitemOpen
  \bibfield  {author} {\bibinfo {author} {\bibfnamefont {H.}~\bibnamefont
  {Casini}}\ and\ \bibinfo {author} {\bibfnamefont {M.}~\bibnamefont
  {Huerta}},\ }\bibfield  {title} {\enquote {\bibinfo {title} {{A c-theorem for
  entanglement entropy}},}\ }\href {\doibase 10.1088/1751-8113/40/25/s57}
  {\bibfield  {journal} {\bibinfo  {journal} {Journal of Physics A:
  Mathematical and Theoretical}\ }\textbf {\bibinfo {volume} {40}},\ \bibinfo
  {pages} {7031--7036} (\bibinfo {year} {2007})}\BibitemShut {NoStop}%
\bibitem [{\citenamefont {Apenko}(2012)}]{Apenko2012}%
  \BibitemOpen
  \bibfield  {author} {\bibinfo {author} {\bibfnamefont {Sergey~M.}\
  \bibnamefont {Apenko}},\ }\bibfield  {title} {\enquote {\bibinfo {title}
  {Information theory and renormalization group flows},}\ }\href {\doibase
  https://doi.org/10.1016/j.physa.2011.08.014} {\bibfield  {journal} {\bibinfo
  {journal} {Physica A: Statistical Mechanics and its Applications}\ }\textbf
  {\bibinfo {volume} {391}},\ \bibinfo {pages} {62 -- 77} (\bibinfo {year}
  {2012})}\BibitemShut {NoStop}%
\bibitem [{\citenamefont {Machta}\ \emph {et~al.}(2013)\citenamefont {Machta},
  \citenamefont {Chachra}, \citenamefont {Transtrum},\ and\ \citenamefont
  {Sethna}}]{Machta604}%
  \BibitemOpen
  \bibfield  {author} {\bibinfo {author} {\bibfnamefont {Benjamin~B.}\
  \bibnamefont {Machta}}, \bibinfo {author} {\bibfnamefont {Ricky}\
  \bibnamefont {Chachra}}, \bibinfo {author} {\bibfnamefont {Mark~K.}\
  \bibnamefont {Transtrum}}, \ and\ \bibinfo {author} {\bibfnamefont
  {James~P.}\ \bibnamefont {Sethna}},\ }\bibfield  {title} {\enquote {\bibinfo
  {title} {{Parameter Space Compression Underlies Emergent Theories and
  Predictive Models}},}\ }\href {\doibase 10.1126/science.1238723} {\bibfield
  {journal} {\bibinfo  {journal} {Science}\ }\textbf {\bibinfo {volume}
  {342}},\ \bibinfo {pages} {604--607} (\bibinfo {year} {2013})}\BibitemShut
  {NoStop}%
\bibitem [{\citenamefont {Balasubramanian}\ \emph {et~al.}(2015)\citenamefont
  {Balasubramanian}, \citenamefont {Heckman},\ and\ \citenamefont
  {Maloney}}]{Balasubramanian:2014bfa}%
  \BibitemOpen
  \bibfield  {author} {\bibinfo {author} {\bibfnamefont {Vijay}\ \bibnamefont
  {Balasubramanian}}, \bibinfo {author} {\bibfnamefont {Jonathan~J.}\
  \bibnamefont {Heckman}}, \ and\ \bibinfo {author} {\bibfnamefont {Alexander}\
  \bibnamefont {Maloney}},\ }\bibfield  {title} {\enquote {\bibinfo {title}
  {{Relative Entropy and Proximity of Quantum Field Theories}},}\ }\href
  {\doibase 10.1007/JHEP05(2015)104} {\bibfield  {journal} {\bibinfo  {journal}
  {JHEP}\ }\textbf {\bibinfo {volume} {05}},\ \bibinfo {pages} {104} (\bibinfo
  {year} {2015})},\ \Eprint {http://arxiv.org/abs/1410.6809} {arXiv:1410.6809
  [hep-th]} \BibitemShut {NoStop}%
\bibitem [{\citenamefont {B{\'e}ny}\ and\ \citenamefont
  {Osborne}(2015{\natexlab{a}})}]{Beny2015a}%
  \BibitemOpen
  \bibfield  {author} {\bibinfo {author} {\bibfnamefont {C{\'e}dric}\
  \bibnamefont {B{\'e}ny}}\ and\ \bibinfo {author} {\bibfnamefont {Tobias~J.}\
  \bibnamefont {Osborne}},\ }\bibfield  {title} {\enquote {\bibinfo {title}
  {The renormalization group via statistical inference},}\ }\href@noop {}
  {\bibfield  {journal} {\bibinfo  {journal} {New Journal of Physics}\ }\textbf
  {\bibinfo {volume} {17}},\ \bibinfo {pages} {083005} (\bibinfo {year}
  {2015}{\natexlab{a}})}\BibitemShut {NoStop}%
\bibitem [{\citenamefont {B{\'e}ny}\ and\ \citenamefont
  {Osborne}(2015{\natexlab{b}})}]{Beny2015b}%
  \BibitemOpen
  \bibfield  {author} {\bibinfo {author} {\bibfnamefont {C{\'e}dric}\
  \bibnamefont {B{\'e}ny}}\ and\ \bibinfo {author} {\bibfnamefont {Tobias~J.}\
  \bibnamefont {Osborne}},\ }\bibfield  {title} {\enquote {\bibinfo {title}
  {Information-geometric approach to the renormalization group},}\ }\href
  {\doibase 10.1103/PhysRevA.92.022330} {\bibfield  {journal} {\bibinfo
  {journal} {Phys. Rev. A}\ }\textbf {\bibinfo {volume} {92}},\ \bibinfo
  {pages} {022330} (\bibinfo {year} {2015}{\natexlab{b}})}\BibitemShut
  {NoStop}%
\bibitem [{\citenamefont {B{\'{e}}ny}(2018)}]{Beny2018coarsegrained}%
  \BibitemOpen
  \bibfield  {author} {\bibinfo {author} {\bibfnamefont {C{\'{e}}dric}\
  \bibnamefont {B{\'{e}}ny}},\ }\bibfield  {title} {\enquote {\bibinfo {title}
  {Coarse-grained distinguishability of field interactions},}\ }\href {\doibase
  10.22331/q-2018-05-24-67} {\bibfield  {journal} {\bibinfo  {journal}
  {{Quantum}}\ }\textbf {\bibinfo {volume} {2}},\ \bibinfo {pages} {67}
  (\bibinfo {year} {2018})}\BibitemShut {NoStop}%
\bibitem [{\citenamefont {Belghazi}\ \emph {et~al.}(2018)\citenamefont
  {Belghazi}, \citenamefont {Baratin}, \citenamefont {Rajeshwar}, \citenamefont
  {Ozair}, \citenamefont {Bengio}, \citenamefont {Courville},\ and\
  \citenamefont {Hjelm}}]{belghazi2018mine}%
  \BibitemOpen
  \bibfield  {author} {\bibinfo {author} {\bibfnamefont {Mohamed~Ishmael}\
  \bibnamefont {Belghazi}}, \bibinfo {author} {\bibfnamefont {Aristide}\
  \bibnamefont {Baratin}}, \bibinfo {author} {\bibfnamefont {Sai}\ \bibnamefont
  {Rajeshwar}}, \bibinfo {author} {\bibfnamefont {Sherjil}\ \bibnamefont
  {Ozair}}, \bibinfo {author} {\bibfnamefont {Yoshua}\ \bibnamefont {Bengio}},
  \bibinfo {author} {\bibfnamefont {Aaron}\ \bibnamefont {Courville}}, \ and\
  \bibinfo {author} {\bibfnamefont {Devon}\ \bibnamefont {Hjelm}},\ }\bibfield
  {title} {\enquote {\bibinfo {title} {{Mutual Information Neural
  Estimation}},}\ \ }(\bibinfo  {publisher} {PMLR},\ \bibinfo {year} {2018})\
  pp.\ \bibinfo {pages} {531--540},\ \Eprint {http://arxiv.org/abs/1801.04062}
  {arXiv:1801.04062} \BibitemShut {NoStop}%
\bibitem [{\citenamefont {Poole}\ \emph {et~al.}(2019)\citenamefont {Poole},
  \citenamefont {Ozair}, \citenamefont {Van Den~Oord}, \citenamefont {Alemi},\
  and\ \citenamefont {Tucker}}]{poole2019variational}%
  \BibitemOpen
  \bibfield  {author} {\bibinfo {author} {\bibfnamefont {Ben}\ \bibnamefont
  {Poole}}, \bibinfo {author} {\bibfnamefont {Sherjil}\ \bibnamefont {Ozair}},
  \bibinfo {author} {\bibfnamefont {Aaron}\ \bibnamefont {Van Den~Oord}},
  \bibinfo {author} {\bibfnamefont {Alex}\ \bibnamefont {Alemi}}, \ and\
  \bibinfo {author} {\bibfnamefont {George}\ \bibnamefont {Tucker}},\
  }\bibfield  {title} {\enquote {\bibinfo {title} {{On Variational Bounds of
  Mutual Information}},}\ \ }(\bibinfo  {publisher} {PMLR},\ \bibinfo {year}
  {2019})\ pp.\ \bibinfo {pages} {5171--5180},\ \Eprint
  {http://arxiv.org/abs/1905.06922} {arXiv:1905.06922} \BibitemShut {NoStop}%
\bibitem [{\citenamefont {G\"okmen}\ \emph {et~al.}()\citenamefont {G\"okmen},
  \citenamefont {Ringel}, \citenamefont {Huber},\ and\ \citenamefont
  {Koch-Janusz}}]{RSMI-NE}%
  \BibitemOpen
  \bibfield  {author} {\bibinfo {author} {\bibfnamefont {Doruk~Efe}\
  \bibnamefont {G\"okmen}}, \bibinfo {author} {\bibfnamefont {Zohar}\
  \bibnamefont {Ringel}}, \bibinfo {author} {\bibfnamefont {Sebastian~D.}\
  \bibnamefont {Huber}}, \ and\ \bibinfo {author} {\bibfnamefont {Maciej}\
  \bibnamefont {Koch-Janusz}},\ }\bibfield  {title} {\enquote {\bibinfo {title}
  {{Real-space mutual information neural estimation (in preparation)}},}\
  }\href@noop {} {\ }\BibitemShut {NoStop}%
\bibitem [{\citenamefont {Koch-Janusz}\ and\ \citenamefont
  {Ringel}(2018)}]{Koch-Janusz2018}%
  \BibitemOpen
  \bibfield  {author} {\bibinfo {author} {\bibfnamefont {Maciej}\ \bibnamefont
  {Koch-Janusz}}\ and\ \bibinfo {author} {\bibfnamefont {Zohar}\ \bibnamefont
  {Ringel}},\ }\bibfield  {title} {\enquote {\bibinfo {title} {Mutual
  information, neural networks and the renormalization group},}\ }\href
  {\doibase 10.1038/s41567-018-0081-4} {\bibfield  {journal} {\bibinfo
  {journal} {Nature Physics}\ }\textbf {\bibinfo {volume} {14}},\ \bibinfo
  {pages} {{578--582}} (\bibinfo {year} {2018})}\BibitemShut {NoStop}%
\bibitem [{\citenamefont {Lenggenhager}\ \emph {et~al.}(2020)\citenamefont
  {Lenggenhager}, \citenamefont {G\"okmen}, \citenamefont {Ringel},
  \citenamefont {Huber},\ and\ \citenamefont
  {Koch-Janusz}}]{PhysRevX.10.011037}%
  \BibitemOpen
  \bibfield  {author} {\bibinfo {author} {\bibfnamefont {Patrick~M.}\
  \bibnamefont {Lenggenhager}}, \bibinfo {author} {\bibfnamefont {Doruk~Efe}\
  \bibnamefont {G\"okmen}}, \bibinfo {author} {\bibfnamefont {Zohar}\
  \bibnamefont {Ringel}}, \bibinfo {author} {\bibfnamefont {Sebastian~D.}\
  \bibnamefont {Huber}}, \ and\ \bibinfo {author} {\bibfnamefont {Maciej}\
  \bibnamefont {Koch-Janusz}},\ }\bibfield  {title} {\enquote {\bibinfo {title}
  {Optimal renormalization group transformation from information theory},}\
  }\href {\doibase 10.1103/PhysRevX.10.011037} {\bibfield  {journal} {\bibinfo
  {journal} {Phys. Rev. X}\ }\textbf {\bibinfo {volume} {10}},\ \bibinfo
  {pages} {011037} (\bibinfo {year} {2020})}\BibitemShut {NoStop}%
\bibitem [{\citenamefont {{Hassanpour}}\ \emph {et~al.}(2017)\citenamefont
  {{Hassanpour}}, \citenamefont {{Wuebben}},\ and\ \citenamefont
  {{Dekorsy}}}]{Hassanpour2017}%
  \BibitemOpen
  \bibfield  {author} {\bibinfo {author} {\bibfnamefont {S.}~\bibnamefont
  {{Hassanpour}}}, \bibinfo {author} {\bibfnamefont {D.}~\bibnamefont
  {{Wuebben}}}, \ and\ \bibinfo {author} {\bibfnamefont {A.}~\bibnamefont
  {{Dekorsy}}},\ }\bibfield  {title} {\enquote {\bibinfo {title} {{Overview and
  Investigation of Algorithms for the Information Bottleneck Method}},}\ }in\
  \href@noop {} {\emph {\bibinfo {booktitle} {SCC 2017; 11th International ITG
  Conference on Systems, Communications and Coding}}}\ (\bibinfo {year}
  {2017})\ pp.\ \bibinfo {pages} {1--6}\BibitemShut {NoStop}%
\bibitem [{\citenamefont {Parker}\ \emph {et~al.}(2002)\citenamefont {Parker},
  \citenamefont {Gedeon},\ and\ \citenamefont {Dimitrov}}]{parkernips}%
  \BibitemOpen
  \bibfield  {author} {\bibinfo {author} {\bibfnamefont {Albert~E.}\
  \bibnamefont {Parker}}, \bibinfo {author} {\bibfnamefont {Tom\'{a}\v{s}}\
  \bibnamefont {Gedeon}}, \ and\ \bibinfo {author} {\bibfnamefont
  {Alexander~G.}\ \bibnamefont {Dimitrov}},\ }\bibfield  {title} {\enquote
  {\bibinfo {title} {{Annealing and the Rate Distortion Problem}},}\ }in\
  \href@noop {} {\emph {\bibinfo {booktitle} {Proceedings of the 15th
  International Conference on Neural Information Processing Systems}}},\
  \bibinfo {series and number} {NIPS’02}\ (\bibinfo  {publisher} {MIT
  Press},\ \bibinfo {address} {Cambridge, MA, USA},\ \bibinfo {year} {2002})\
  p.\ \bibinfo {pages} {993–976}\BibitemShut {NoStop}%
\bibitem [{\citenamefont {Gedeon}\ \emph {et~al.}(2012)\citenamefont {Gedeon},
  \citenamefont {Parker},\ and\ \citenamefont {Dimitrov}}]{parkerentropy}%
  \BibitemOpen
  \bibfield  {author} {\bibinfo {author} {\bibfnamefont {Tomas}\ \bibnamefont
  {Gedeon}}, \bibinfo {author} {\bibfnamefont {Albert~E.}\ \bibnamefont
  {Parker}}, \ and\ \bibinfo {author} {\bibfnamefont {Alexander.G}\
  \bibnamefont {Dimitrov}},\ }\bibfield  {title} {\enquote {\bibinfo {title}
  {{The Mathematical structure of Information Bottleneck Methods}},}\
  }\href@noop {} {\bibfield  {journal} {\bibinfo  {journal} {Entropy}\ }\textbf
  {\bibinfo {volume} {14}},\ \bibinfo {pages} {456--479} (\bibinfo {year}
  {2012})}\BibitemShut {NoStop}%
\bibitem [{\citenamefont {Chechik}\ \emph {et~al.}(2004)\citenamefont
  {Chechik}, \citenamefont {Globerson}, \citenamefont {Tishby},\ and\
  \citenamefont {Weiss}}]{NIPS2003_2457}%
  \BibitemOpen
  \bibfield  {author} {\bibinfo {author} {\bibfnamefont {Gal}\ \bibnamefont
  {Chechik}}, \bibinfo {author} {\bibfnamefont {Amir}\ \bibnamefont
  {Globerson}}, \bibinfo {author} {\bibfnamefont {Naftali}\ \bibnamefont
  {Tishby}}, \ and\ \bibinfo {author} {\bibfnamefont {Yair}\ \bibnamefont
  {Weiss}},\ }\bibfield  {title} {\enquote {\bibinfo {title} {{Information
  Bottleneck for Gaussian Variables}},}\ }in\ \href@noop {} {\emph {\bibinfo
  {booktitle} {Advances in Neural Information Processing Systems 16}}},\
  \bibinfo {editor} {edited by\ \bibinfo {editor} {\bibfnamefont
  {S.}~\bibnamefont {Thrun}}, \bibinfo {editor} {\bibfnamefont {L.~K.}\
  \bibnamefont {Saul}}, \ and\ \bibinfo {editor} {\bibfnamefont
  {B.}~\bibnamefont {Sch\"{o}lkopf}}}\ (\bibinfo  {publisher} {MIT Press},\
  \bibinfo {year} {2004})\ pp.\ \bibinfo {pages} {1213--1220}\BibitemShut
  {NoStop}%
\bibitem [{\citenamefont {{Kramers}}\ and\ \citenamefont
  {{Wannier}}(1941)}]{1941PhRv...60..252K}%
  \BibitemOpen
  \bibfield  {author} {\bibinfo {author} {\bibfnamefont {H.~A.}\ \bibnamefont
  {{Kramers}}}\ and\ \bibinfo {author} {\bibfnamefont {G.~H.}\ \bibnamefont
  {{Wannier}}},\ }\bibfield  {title} {\enquote {\bibinfo {title} {{Statistics
  of the Two-Dimensional Ferromagnet. Part I}},}\ }\href {\doibase
  10.1103/PhysRev.60.252} {\bibfield  {journal} {\bibinfo  {journal} {Physical
  Review}\ }\textbf {\bibinfo {volume} {60}},\ \bibinfo {pages} {252--262}
  (\bibinfo {year} {1941})}\BibitemShut {NoStop}%
\bibitem [{\citenamefont {{Onsager}}(1944)}]{1944PhRv...65..117O}%
  \BibitemOpen
  \bibfield  {author} {\bibinfo {author} {\bibfnamefont {Lars}\ \bibnamefont
  {{Onsager}}},\ }\bibfield  {title} {\enquote {\bibinfo {title} {{Crystal
  Statistics. I. A Two-Dimensional Model with an Order-Disorder Transition}},}\
  }\href {\doibase 10.1103/PhysRev.65.117} {\bibfield  {journal} {\bibinfo
  {journal} {Physical Review}\ }\textbf {\bibinfo {volume} {65}},\ \bibinfo
  {pages} {117--149} (\bibinfo {year} {1944})}\BibitemShut {NoStop}%
\bibitem [{\citenamefont {Nightingale}(1982)}]{doi:10.1063/1.330232}%
  \BibitemOpen
  \bibfield  {author} {\bibinfo {author} {\bibfnamefont {Peter}\ \bibnamefont
  {Nightingale}},\ }\bibfield  {title} {\enquote {\bibinfo {title}
  {{Finite-size scaling and phenomenological renormalization}},}\ }\href
  {\doibase 10.1063/1.330232} {\bibfield  {journal} {\bibinfo  {journal}
  {Journal of Applied Physics}\ }\textbf {\bibinfo {volume} {53}},\ \bibinfo
  {pages} {7927--7932} (\bibinfo {year} {1982})},\ \Eprint
  {http://arxiv.org/abs/https://doi.org/10.1063/1.330232}
  {https://doi.org/10.1063/1.330232} \BibitemShut {NoStop}%
\bibitem [{\citenamefont {Derrida}\ and\ \citenamefont
  {De~Seze}(1982)}]{latticeanimals}%
  \BibitemOpen
  \bibfield  {author} {\bibinfo {author} {\bibfnamefont {B.}~\bibnamefont
  {Derrida}}\ and\ \bibinfo {author} {\bibfnamefont {L.}~\bibnamefont
  {De~Seze}},\ }\bibfield  {title} {\enquote {\bibinfo {title} {{Application of
  the phenomenological renormalization to percolation and lattice animals in
  dimension 2}},}\ }\href@noop {} {\bibfield  {journal} {\bibinfo  {journal}
  {J.Physique}\ }\textbf {\bibinfo {volume} {43}},\ \bibinfo {pages} {475 --
  483} (\bibinfo {year} {1982})}\BibitemShut {NoStop}%
\bibitem [{\citenamefont {Cardy}(1984)}]{Cardy_1984}%
  \BibitemOpen
  \bibfield  {author} {\bibinfo {author} {\bibfnamefont {J~L}\ \bibnamefont
  {Cardy}},\ }\bibfield  {title} {\enquote {\bibinfo {title} {{Conformal
  invariance and universality in finite-size scaling}},}\ }\href {\doibase
  10.1088/0305-4470/17/7/003} {\bibfield  {journal} {\bibinfo  {journal}
  {Journal of Physics A: Mathematical and General}\ }\textbf {\bibinfo {volume}
  {17}},\ \bibinfo {pages} {L385--L387} (\bibinfo {year} {1984})}\BibitemShut
  {NoStop}%
\bibitem [{\citenamefont {Cardy}(1986)}]{CARDY1986186}%
  \BibitemOpen
  \bibfield  {author} {\bibinfo {author} {\bibfnamefont {John~L.}\ \bibnamefont
  {Cardy}},\ }\bibfield  {title} {\enquote {\bibinfo {title} {Operator content
  of two-dimensional conformally invariant theories},}\ }\href {\doibase
  https://doi.org/10.1016/0550-3213(86)90552-3} {\bibfield  {journal} {\bibinfo
   {journal} {Nuclear Physics B}\ }\textbf {\bibinfo {volume} {270}},\ \bibinfo
  {pages} {186 -- 204} (\bibinfo {year} {1986})}\BibitemShut {NoStop}%
\bibitem [{\citenamefont {Banerjee}\ and\ \citenamefont {Ringel}()}]{aditya}%
  \BibitemOpen
  \bibfield  {author} {\bibinfo {author} {\bibfnamefont {Aditya}\ \bibnamefont
  {Banerjee}}\ and\ \bibinfo {author} {\bibfnamefont {Zohar}\ \bibnamefont
  {Ringel}},\ }\bibfield  {title} {\enquote {\bibinfo {title} {{Information
  bottleneck and Gaussian field theory (in preparation).}}}\ }\href@noop {} {\
  }\BibitemShut {NoStop}%
\bibitem [{\citenamefont {Zinn-Justin}(1989)}]{zinn1989quantum}%
  \BibitemOpen
  \bibfield  {author} {\bibinfo {author} {\bibfnamefont {J.}~\bibnamefont
  {Zinn-Justin}},\ }\href@noop {} {\emph {\bibinfo {title} {{Quantum Field
  Theory and Critical Phenomena}}}},\ International series of monographs on
  physics\ (\bibinfo  {publisher} {Clarendon Press},\ \bibinfo {year}
  {1989})\BibitemShut {NoStop}%
\bibitem [{\citenamefont {Yang}\ and\ \citenamefont {Zhang}(1990)}]{Zhang1990}%
  \BibitemOpen
  \bibfield  {author} {\bibinfo {author} {\bibfnamefont {Chen~Ning}\
  \bibnamefont {Yang}}\ and\ \bibinfo {author} {\bibfnamefont {S.C.}\
  \bibnamefont {Zhang}},\ }\bibfield  {title} {\enquote {\bibinfo {title} {{SO4
  Symmetry in a Hubbard model}},}\ }\href {\doibase 10.1142/S0217984990000933}
  {\bibfield  {journal} {\bibinfo  {journal} {Modern Physics Letters B}\
  }\textbf {\bibinfo {volume} {04}},\ \bibinfo {pages} {759--766} (\bibinfo
  {year} {1990})}\BibitemShut {NoStop}%
\bibitem [{\citenamefont {Senthil}\ \emph {et~al.}(2004)\citenamefont
  {Senthil}, \citenamefont {Vishwanath}, \citenamefont {Balents}, \citenamefont
  {Sachdev},\ and\ \citenamefont {Fisher}}]{Senthil2004}%
  \BibitemOpen
  \bibfield  {author} {\bibinfo {author} {\bibfnamefont {T.}~\bibnamefont
  {Senthil}}, \bibinfo {author} {\bibfnamefont {Ashvin}\ \bibnamefont
  {Vishwanath}}, \bibinfo {author} {\bibfnamefont {Leon}\ \bibnamefont
  {Balents}}, \bibinfo {author} {\bibfnamefont {Subir}\ \bibnamefont
  {Sachdev}}, \ and\ \bibinfo {author} {\bibfnamefont {Matthew P.~A.}\
  \bibnamefont {Fisher}},\ }\bibfield  {title} {\enquote {\bibinfo {title}
  {{Deconfined Quantum Critical Points}},}\ }\href {\doibase
  10.1126/science.1091806} {\bibfield  {journal} {\bibinfo  {journal}
  {Science}\ }\textbf {\bibinfo {volume} {303}},\ \bibinfo {pages} {1490--1494}
  (\bibinfo {year} {2004})}\BibitemShut {NoStop}%
\bibitem [{\citenamefont {Bondesan}\ and\ \citenamefont
  {Lamacraft}(2019)}]{Bondesan2019}%
  \BibitemOpen
  \bibfield  {author} {\bibinfo {author} {\bibfnamefont {Roberto}\ \bibnamefont
  {Bondesan}}\ and\ \bibinfo {author} {\bibfnamefont {Austen}\ \bibnamefont
  {Lamacraft}},\ }\bibfield  {title} {\enquote {\bibinfo {title} {{Learning
  Symmetries of Classical Integrable Systems}},}\ }\href
  {http://arxiv.org/abs/1906.04645} {\ \textbf {\bibinfo {volume}
  {abs/1906.04645}} (\bibinfo {year} {2019})},\ \Eprint
  {http://arxiv.org/abs/1906.04645} {arXiv:1906.04645} \BibitemShut {NoStop}%
\bibitem [{\citenamefont {Schneidman}\ \emph {et~al.}(2001)\citenamefont
  {Schneidman}, \citenamefont {Slonim}, \citenamefont {Tishby}, \citenamefont
  {de~Ruyter~van Steveninck},\ and\ \citenamefont {Bialek}}]{Schneidman}%
  \BibitemOpen
  \bibfield  {author} {\bibinfo {author} {\bibfnamefont {Elad}\ \bibnamefont
  {Schneidman}}, \bibinfo {author} {\bibfnamefont {Noam}\ \bibnamefont
  {Slonim}}, \bibinfo {author} {\bibfnamefont {Naftali}\ \bibnamefont
  {Tishby}}, \bibinfo {author} {\bibfnamefont {Rob~R.}\ \bibnamefont
  {de~Ruyter~van Steveninck}}, \ and\ \bibinfo {author} {\bibfnamefont
  {William}\ \bibnamefont {Bialek}},\ }\href@noop {} {\enquote {\bibinfo
  {title} {{Analyzing Neural Codes Using the Information Bottleneck Method}},}\
  } (\bibinfo {year} {2001})\BibitemShut {NoStop}%
\bibitem [{\citenamefont {Creutzig}\ and\ \citenamefont
  {Sprekeler}(2008)}]{IBslowness}%
  \BibitemOpen
  \bibfield  {author} {\bibinfo {author} {\bibfnamefont {Felix}\ \bibnamefont
  {Creutzig}}\ and\ \bibinfo {author} {\bibfnamefont {Henning}\ \bibnamefont
  {Sprekeler}},\ }\bibfield  {title} {\enquote {\bibinfo {title} {{Predictive
  Coding and the Slowness Principle: An Information-Theoretic Approach}},}\
  }\href {\doibase 10.1162/neco.2008.01-07-455} {\bibfield  {journal} {\bibinfo
   {journal} {Neural Computation}\ }\textbf {\bibinfo {volume} {20}},\ \bibinfo
  {pages} {1026--1041} (\bibinfo {year} {2008})}\BibitemShut {NoStop}%
\bibitem [{\citenamefont {Buesing}\ and\ \citenamefont
  {Maass}(2010)}]{IBspiking}%
  \BibitemOpen
  \bibfield  {author} {\bibinfo {author} {\bibfnamefont {Lars}\ \bibnamefont
  {Buesing}}\ and\ \bibinfo {author} {\bibfnamefont {Wolfgang}\ \bibnamefont
  {Maass}},\ }\bibfield  {title} {\enquote {\bibinfo {title} {{A Spiking Neuron
  as Information Bottleneck}},}\ }\href {\doibase 10.1162/neco.2010.08-09-1084}
  {\bibfield  {journal} {\bibinfo  {journal} {Neural Computation}\ }\textbf
  {\bibinfo {volume} {22}},\ \bibinfo {pages} {1961--1992} (\bibinfo {year}
  {2010})},\ \bibinfo {note} {pMID: 20337537}\BibitemShut {NoStop}%
\bibitem [{\citenamefont {Slonim}\ and\ \citenamefont
  {Tishby}(2000)}]{10.1145/345508.345578}%
  \BibitemOpen
  \bibfield  {author} {\bibinfo {author} {\bibfnamefont {Noam}\ \bibnamefont
  {Slonim}}\ and\ \bibinfo {author} {\bibfnamefont {Naftali}\ \bibnamefont
  {Tishby}},\ }\bibfield  {title} {\enquote {\bibinfo {title} {{Document
  Clustering Using Word Clusters via the Information Bottleneck Method}},}\
  }in\ \href {\doibase 10.1145/345508.345578} {\emph {\bibinfo {booktitle}
  {Proceedings of the 23rd Annual International ACM SIGIR Conference on
  Research and Development in Information Retrieval}}},\ \bibinfo {series and
  number} {SIGIR '00}\ (\bibinfo {year} {2000})\ p.\ \bibinfo {pages}
  {208–215}\BibitemShut {NoStop}%
\bibitem [{\citenamefont {Still}\ \emph {et~al.}(2004)\citenamefont {Still},
  \citenamefont {Bialek},\ and\ \citenamefont {Bottou}}]{NIPS2003_794288f2}%
  \BibitemOpen
  \bibfield  {author} {\bibinfo {author} {\bibfnamefont {Susanne}\ \bibnamefont
  {Still}}, \bibinfo {author} {\bibfnamefont {William}\ \bibnamefont {Bialek}},
  \ and\ \bibinfo {author} {\bibfnamefont {L\'{e}on}\ \bibnamefont {Bottou}},\
  }\bibfield  {title} {\enquote {\bibinfo {title} {{Geometric Clustering Using
  the Information Bottleneck Method}},}\ }in\ \href
  {https://proceedings.neurips.cc/paper/2003/file/794288f252f45d35735a13853e605939-Paper.pdf}
  {\emph {\bibinfo {booktitle} {Advances in Neural Information Processing
  Systems}}},\ Vol.~\bibinfo {volume} {16},\ \bibinfo {editor} {edited by\
  \bibinfo {editor} {\bibfnamefont {S.}~\bibnamefont {Thrun}}, \bibinfo
  {editor} {\bibfnamefont {L.}~\bibnamefont {Saul}}, \ and\ \bibinfo {editor}
  {\bibfnamefont {B.}~\bibnamefont {Sch\"{o}lkopf}}}\ (\bibinfo  {publisher}
  {MIT Press},\ \bibinfo {year} {2004})\ pp.\ \bibinfo {pages}
  {1165--1172}\BibitemShut {NoStop}%
\bibitem [{\citenamefont {Strouse}\ and\ \citenamefont
  {Schwab}(2019)}]{IBgeometriccluster}%
  \BibitemOpen
  \bibfield  {author} {\bibinfo {author} {\bibfnamefont {DJ}~\bibnamefont
  {Strouse}}\ and\ \bibinfo {author} {\bibfnamefont {David~J.}\ \bibnamefont
  {Schwab}},\ }\bibfield  {title} {\enquote {\bibinfo {title} {{The Information
  Bottleneck and Geometric Clustering}},}\ }\href {\doibase
  10.1162/neco\_a\_01136} {\bibfield  {journal} {\bibinfo  {journal} {Neural
  Computation}\ }\textbf {\bibinfo {volume} {31}},\ \bibinfo {pages} {596--612}
  (\bibinfo {year} {2019})},\ \bibinfo {note} {pMID: 30314426}\BibitemShut
  {NoStop}%
\bibitem [{\citenamefont {Creutzig}\ \emph {et~al.}(2009)\citenamefont
  {Creutzig}, \citenamefont {Globerson},\ and\ \citenamefont
  {Tishby}}]{PhysRevE.79.041925}%
  \BibitemOpen
  \bibfield  {author} {\bibinfo {author} {\bibfnamefont {Felix}\ \bibnamefont
  {Creutzig}}, \bibinfo {author} {\bibfnamefont {Amir}\ \bibnamefont
  {Globerson}}, \ and\ \bibinfo {author} {\bibfnamefont {Naftali}\ \bibnamefont
  {Tishby}},\ }\bibfield  {title} {\enquote {\bibinfo {title} {{Past-future
  information bottleneck in dynamical systems}},}\ }\href {\doibase
  10.1103/PhysRevE.79.041925} {\bibfield  {journal} {\bibinfo  {journal} {Phys.
  Rev. E}\ }\textbf {\bibinfo {volume} {79}},\ \bibinfo {pages} {041925}
  (\bibinfo {year} {2009})}\BibitemShut {NoStop}%
\bibitem [{\citenamefont {Still}(2014)}]{e16020968}%
  \BibitemOpen
  \bibfield  {author} {\bibinfo {author} {\bibfnamefont {Susanne}\ \bibnamefont
  {Still}},\ }\bibfield  {title} {\enquote {\bibinfo {title} {{Information
  Bottleneck Approach to Predictive Inference}},}\ }\href {\doibase
  10.3390/e16020968} {\bibfield  {journal} {\bibinfo  {journal} {Entropy}\
  }\textbf {\bibinfo {volume} {16}},\ \bibinfo {pages} {968--989} (\bibinfo
  {year} {2014})}\BibitemShut {NoStop}%
\bibitem [{\citenamefont {Benger}\ and\ \citenamefont {Tishby}()}]{etam}%
  \BibitemOpen
  \bibfield  {author} {\bibinfo {author} {\bibfnamefont {Etam}\ \bibnamefont
  {Benger}}\ and\ \bibinfo {author} {\bibfnamefont {Naftali}\ \bibnamefont
  {Tishby}},\ }\bibfield  {title} {\enquote {\bibinfo {title} {{All Information
  Bottleneck Phase Transitions are Local (in preparation).}}}\ }\href@noop {}
  {\ }\BibitemShut {NoStop}%
\bibitem [{\citenamefont {Slonim}(2002)}]{slonim2002information}%
  \BibitemOpen
  \bibfield  {author} {\bibinfo {author} {\bibfnamefont {Noam}\ \bibnamefont
  {Slonim}},\ }\emph {\bibinfo {title} {{The information bottleneck: Theory and
  applications}}},\ \href@noop {} {Ph.D. thesis} (\bibinfo {year}
  {2002})\BibitemShut {NoStop}%
\end{thebibliography}%

\appendix

\section{The IB Equations}

The IB problem Ref.\cite{infbottle1}, as described in the main text, is set up as a minimization problem over the class of conditional probability distributions $P(h|v)$ of the following IB Lagrangian:
\begin{equation}\label{eq:app_ib_lag1}
\mathcal{L}_{IB}[P(H|V)] \equiv I(V;H) - \beta I(H;E),
\end{equation}
Somewhat surprisingly (since the terms in $\mathcal{L}_{IB}$ are highly nonlinear) a formal solution can be found by performing the variation $\delta \mathcal{L}_{IB} / \delta P(h|v) =0$. As shown in Ref.\cite{infbottle1}, the optimal solution can be written as:
\begin{align}\label{eq:app_infbott_1}
P(h|v) = \frac{P(h)}{Z(v,\beta)} \mathrm{exp}\left( -\beta \mathrm{D}_{KL}[P(e|x)|P(e|h)] \right),
\end{align}
where $\mathrm{D}_{KL}$ is the Kullback-Leibler divergence of the conditional probability distributions:
\begin{align}\label{eq:KL}
\mathrm{D}_{KL}[P(e|x)|P(e|h)] = \sum_{e} P(e|v)\log( \frac{P(e|v)}{P(e|h)}),
\end{align}
$Z$ is a normalizing factor and:
\begin{align}\label{eq:app_infbott_2}
P(e|h) = \frac{1}{P(h)}\sum_v P(e|v)P(h|v)P(v),
\end{align}

Note that this is only a formal solution, which is in fact implicit. It does, however, reveal that the optimal encoder is one which results in the minimal distortion, as measured by $\mathrm{D}_{KL}$, of recovery of $e$ when using the compressed variable $h$ in place of the original $v$.

The simplest way to find the solutions explicitly is to convert the self-consistent Eqs.\ref{eq:app_infbott_1} and \ref{eq:app_infbott_2} into an iterative algorithm, which can be shown to converge Ref.\cite{infbottle1}. These are the explicit IB equations, whose shortened form we write out in the main text as Eqs.\ref{eq:ibeqs1_maintext}:
\begin{align}\label{eq:app_ib_it1}
P(h|v) &= \frac{P(h)}{Z} \mathrm{exp}\left(-\beta \sum_{e} P(e|v)\log\left[\frac{P(e|v)}{P(e|h)}\right]\right) \\\label{eq:app_ib_it2}
P(h) &= \sum_{v} P(h|v)P(v)\\\label{eq:app_ib_it3}
P(e|h) &= \sum_{v} P(e|v)P(v|h),
\end{align} 
where $Z = \sum_{h} P(h) \mathrm{exp}( -\beta \sum_{e} P(e|v)\log( \frac{P(e|v)}{P(e|h)}))$. 

In the short-hand form of the IB equations written in the main text, Eqs.\ref{eq:ibeqs1_maintext}, the equation \ref{eq:app_ib_it2} was not written out explicitly, and equation \ref{eq:app_ib_it1} was slightly massaged:
\begin{equation} \nonumber
\begin{split}
P(h|v) &= \frac{P(h) e^{-\beta \sum_{e} P(e|v)\log(\frac{P(e|v)}{P(e|h)})}}{\sum_{h} P(h) e^{-\beta \sum_{e} P(e|v)\log( \frac{P(e|v)}{P(e|h)})}}\\
&=\frac{P(h) e^{\beta \sum_{e} P(e|v)\log(P(e|h))}}{\sum_{h} P(h) e^{\beta \sum_{e} P(e|v)\log(P(e|h))}}
\end{split}
\end{equation}
The denominator is $h$-independent, and the equation was given up to a proportionality constant in the main text.

The iterative IB algorithm, while useful in the theoretical investigations (and also used in the small numerical validation experiment in the appendix below) is not the only, nor necessarily the best numerical technique to solve the IB equations. For an overview of other methods we refer to Ref.\cite{Hassanpour2017}. Nevertheless, directly solving the IB equations for larger input distributions is generally computationally hard. Recently, an entirely different approach was developed in the context of deep learning \cite{AlemiFD016}. Instead of solving the IB equations, the IB Lagrangian, or a bound on it, is taken as a cost function, and the optimal encoder is parametrised by a deep neural network and optimized using \emph{e.g.}~stochastic gradient descent. This allows to exploit the numerical efficiency of machine learning toolboxes.
A similar technique is used for the optimization of $I(H;E)$ in RSMI-NE \cite{RSMI-NE}.

We remark here that while IB is phrased as an abstract compression theory problem, it has found applications in computational neuroscience, where the question of what is the fundamentally important information extracted from say, neuronal activity measurements, is nontrivial \cite{Schneidman,IBslowness,IBspiking}. Furthermore, IB has been used in computer science problems, \emph{e.g.}~clustering analyses \cite{10.1145/345508.345578,NIPS2003_794288f2,IBgeometriccluster}, but also in attempts to quantify relevant or predictive information in physics, mostly in the context of temporal correlations in non-equilibrium systems, see \emph{e.g}~\cite{PhysRevE.79.041925,e16020968}.

\section{From the transfer matrix to the reduced IB equation}

In this appendix we connect the transfer matrix (TM) viewpoint to the conditional probabilities used in the main text, and derive the reduced IB equations.

In any local lattice model on the (hyper-) cylinder the partition function $\mathcal{Z}$ can be written in terms of the trace of the transfer matrix $\mathcal{T}$ as $\mathcal{Z} = tr(\mathcal{T}^{L_\infty})$, where $L_\infty$ is the length of the system (here with periodic boundary conditions). For a system described by a conformal field theory (CFT), the eigenvectors and eigenvalues of $\mathcal{T}$ correspond to operators in the CFT \cite{Cardy_1984,CARDY1986186}. 
For simplicity we consider conformal theories which have a single most relevant operator, and a corresponding microscopic lattice model with finite-range interactions. Assuming the definitions of IB quantities given in the main text, we will show that in the large buffer limit: \textbf{(i)} The description of the coarse-graining cell $V$ and environment cell $E$ can be reduced to only two random variables associated with certain ``weak-expectation-values" $r_e$ and $r_v$ of most relevant primary operator on the boundary of these two regions. \textbf{(ii)} All marginal and conditional probability distributions relevant for IB can be expressed in terms of these two random variables. \textbf{(iii)} The resulting IB equations can be solved close to the first critical value of $\beta_{c,1}$, yielding the optimal encoding which, for $\beta = \beta_{c,1} + t$, amounts to tracking the above weak-expectation-values of the physically most relevant operator (which are the extracted ``features", in machine learning parlance). 

Consider then a statistical-mechanical system on an infinite cylinder, with a finite cylindrical coarse-graining cell $V$ and environment $E$ (playing the role of the ``relevance" variable) to its right, separated by a buffer, as depicted in Fig.\ref{fig:transfer2} in the main text. We assume a microscopic structure, for concreteness we take a square lattice. The transfer matrix $\mathcal{T}$, as usual,  acts on the elementary slices of the cylinder, each consisting of a single (periodic) row of lattice sites.
We denote by $L$ the number of sites on the circumference of the cylinder and by $L_B$ length of the buffer. We further denote by $\partial V_R$ ($\partial E_L$) the configuration of degrees of freedom on the right-most (left-most) slice of sites in $V$ ($E$). These can be thought of as basis vectors of the vector space on which the transfer matrix acts, and so we shall denote them by $\langle \partial X |$ or $|\partial X \rangle = [\langle \partial X |]^{\mathrm{T}}$, depending on whether $\mathcal{T}$ acts on them from the left or from the right ($X = V$ or $E$ as applicable).

It is well known \cite{Cardy_1984,CARDY1986186} that the eigenvalues $\lambda_i$ and eigenvectors $|i\rangle$ of the transfer matrix have a direct relation to the CFT's operator content. Namely, for a square lattice, $\lambda_i/\lambda_0 = e^{-\frac{2\pi}{L}\Delta_i}$, where $\Delta_i$ is the total (\emph{i.e.}~sum of the holomorphic and antiholomorphic) scaling dimension. We shall from now on normalize so that $\lambda_0=1$ in the limit of large circumference $L$. Here we consider the case where $\Delta_2 >\Delta_1 > 0$, and we take the buffer to be much larger than ratio $\Delta_1/L$. We will see below that in this limit, the relevant degrees of freedom, or in other words the relevant random variables, in ${\rm V}$ and ${\rm E}$ are: 
\begin{align}\label{eq:rxre1}
r_v &= \frac{\langle 1 |\partial V_R \rangle}{\langle 0 | \partial V_R \rangle} = \frac{\langle 0| \phi_{\Delta_1} |\partial V_R \rangle}{\langle 0 | \partial V_R \rangle} \\ \nonumber
r_e &= \frac{\langle \partial E_L |1 \rangle}{\langle \partial E_L | 0 \rangle} = \frac{\langle \partial E_L|\phi_{\Delta_1}  |0 \rangle}{\langle \partial E_L | 0 \rangle}
\end{align}
where on both right-hand-sides we used the operator state correspondence relating the action of a primary operator $\phi_{\Delta_1} $ on the identity state $|0\rangle$ with the state $|i\rangle$. 

In a quantum mechanical setting such operator expectation values as those appearing on the r.h.s. go under the name weak-values. In cases where the primary operator turns out to be diagonal in the transfer matrix basis, the $\langle 0 | \partial V_R\rangle$ and $\langle  \partial E_L | 0 \rangle $ factors cancel and $r_{v/e}$ simply becomes the diagonal elements of that operator. As an example, for the Ising model $r_v$ is the overall magnetization on $\partial V$. We note by passing that for a compact boson $\phi \in [0,R)$, $r_{v/e}$ would be the two leading electric vertex operators \cite{aditya} associated with the zero transverse-momentum component of $\phi(x)$.

{\bf The conditional probabilities of the system.} Consider, for concreteness, the conditional probability $P(v|e)$, which enters the IB equations. We assume an infinite cylinder to the left of ${\rm V}$ and right of ${\rm E}$, and work in the limit of large buffer $L_B$. With simple transfer matrix manipulations (using TM representation of probability distributions similar to the one below Eq.\ref{eq:ibeqs1_maintext} in the main text, and the eigendecompostion $\mathcal{T} = |0\rangle\langle 0| + \sum_{i} e^{-2\pi \Delta_i /L} |\Delta_i\rangle\langle \Delta_i|$), it can be written as:
\begin{align}
\nonumber P(v|e) & = \frac{1}{N} \langle 0 | \partial V_L \rangle P_{free BC}({\rm V})\left[ \langle \partial V_R|1\rangle\langle 1| \partial E_L \rangle \lambda_1^{L_B} \right. \\ \label{eq:pxe1}
&+\left. \langle \partial V_R|0\rangle\langle 0| \partial E_L \rangle\lambda_0^{L_B}\right]  + \mathcal{O}\left( (\lambda_2 / \lambda_0 )^{L_B}\right)
\end{align}
Here $\partial V_L$ denotes the configuration on the left boundary of $V$, $P_{freeBC}(V)$ is the probability of the sub-system $V$ with free boundary conditions, which is related to the marginal probability of the sub-system $V$ via $P(v) = N^{-1} \langle 0 | \partial V_L\rangle \langle\partial V_R| 0 \rangle P_{freeBC}({\rm V})$, and $ N$ is the normalization factor. In what follows we will drop the exponentially suppressed terms of order $\mathcal{O}\left( (\lambda_2 / \lambda_0 )^{L_B}\right)$. Explicitly written, $P_{freeBC}(V)$ is given by:
\begin{equation}
P_{freeBC}(V) \propto \langle \partial V_L| \mathcal{T}|x_2\rangle\langle x_2 | \mathcal{T}\ldots \mathcal{T}|\partial V_R\rangle
\end{equation}
That is, it is simply the cumulative action of the transfer matrix on the slices of $V$. Note that the term in the square brackets in Eq.\ref{eq:pxe1} comes from the action of the transfer matrix along the buffer, starting from the right boundary of $V$ and ending at the left boundary of $E$. Taking out a factor of $P(V)$ and absorbing all normalization factors to a factor $N$, one obtains:  
\begin{align}\nonumber
P(v|e) &= N^{-1} P(v)\left[ 1 + \frac{\langle \partial V_R|1\rangle \langle 1| \partial E_L\rangle}{\langle \partial V_R|0\rangle \langle 0| \partial E_L \rangle}\left(\frac{\lambda_1}{\lambda_0}\right)^{L_B}\right]  \\ \label{eq:pxe2}
&= N^{-1} P(v)\left[ 1 + \epsilon r_e r_v \right],
\end{align}
where $\epsilon = \left(\lambda_1 / \lambda_0\right)^{L_B}$. The normalization $N$ is given by:
\begin{align}\label{eq:norm1}
\frac{1}{1+\langle r_v\rangle r_e \epsilon},
\end{align}
with:
\begin{align}\nonumber
\langle r_v \rangle &= \sum_{v} P(v) r_v = \sum_{\partial V_R} P(\partial V_R) r_v  \\ \label{eq:norm2}
&= \sum_{\partial V_R}\langle 0 | \partial V_R \rangle \langle \partial V_R |0\rangle \frac{\langle 1 |\partial V_R \rangle}{\langle 0 | \partial V_R \rangle} = \langle 1 | 0 \rangle = 0
\end{align}
The summation is over all configurations of $\partial V_R$. Following this we find that $\langle r_v \rangle =\langle r_e \rangle = 0$ and therefore $N=1$. Thus, as advertised, ${\rm V}$ depends on ${\rm E}$ only through $r_e$ \emph{i.e.}~$P(v|e)=P(v|r_e)$, and the same holds for $P(e|v)$. 
One can also show that the variances of $r_{e/v}$ obey $\langle r_e^2 \rangle = \langle r_v^2 \rangle = 1$. 

{\bf The IB equation.} We wish to solve the IB equation for the optimal encoder $P(h|v)$ given by \cite{infbottle1}:  
\begin{align}\label{eq:ibeqs1}
P(h|v) &\propto P(h) e^{\beta \sum_{e} P(e|v)\log( P(e|h))} \\ \nonumber 
P(e|h) &= \sum_{v} P(e|v)P(v|h)
\end{align}
Here, as before, $e,h$ and $v$ denote configurations of $E, H, V$ respectively, and the symbol $\propto$ means up to an $h$-independent normalization factor. In order to find the solution we next establish, generalizing the computations above, that the conditional probabilities $P(v|e),P(e|v)$ and the ``decoder" $P(e|h)$ depend on each other 
only through $r_v$ and $r_e$. 

{\bf The reduced IB equation.} The IB equation \ref{eq:ibeqs1} for the encoder is difficult to solve, since it involves a summation over the entire configuration space of $E$ and, furthermore, it is coupled to the equation for the decoder which involves a summation over all configurations of $V$. It is therefore highly beneficial to reduce these equation to ones involving only the configuration space of $r_e$ and $r_v$. To this end we first note that the dependence of the encoder on $V$ in the IB equation only appears through $P(e|v)$, and therefore can be replaced by $r_v$. Similarly we find: 
\begin{align}\label{eq:phrx1}
P(h|r_v) &\propto P(h) e^{\beta \sum_{e} P(e|r_v)\log\left(\sum_{v'}P(e|r_{v'})P(v'|h)\right)}. 
\end{align}
Next we rewrite $P(e|r_{v'})=P(e)[1+\epsilon r_v r_e]$ and expand to first order in $\epsilon$ to obtain the reduced IB equation:
\begin{align}
\label{Eq:reducedIB}
P(h|r_v) &\propto P(h) e^{\beta \epsilon^2 \langle r_e^2\rangle r_v \langle r_v \rangle_{h}} = P(h) e^{\beta \epsilon^2 r_v \langle r_v \rangle_{h}}
\end{align}
where $\langle r_v \rangle_{h}$ is the expectation value of $r_v$ given $h$, based on the joint probability $P(h,v,e) = P(v,e)P(h|v)$.

Equation \ref{Eq:reducedIB} is the key results of this section. It shows that the optimal encoder depends on V only through $r_v$ and in the above specific exponential manner. It changes for one CFT to another through possible values of $r_v$ and the conditional expectation value $\langle r_v \rangle_{h}$. As a sanity check one finds that $P(h|r_v) = const.$ is always a solution. At small enough $\beta$ it is the only one, above the IB phase transitions it is an unstable, suboptimal one (see below). Observe that Eq.\ref{Eq:reducedIB}, though simplified, is still an implicit equation, as the quantities in the exponent depend on the left-hand side. It can, however, be solved explicitly in the vicinity of the first phase transition (which corresponds to the encoder beginning to track the first, most relevant, feature of the data).

\section{Solving the reduced IB equation}
Following \cite{parkerentropy} we consider the encoder at $\beta = \beta_{c,1} + t$ where $\beta_{c,1}$ marks the first breaking of the permutation symmetry of the encoder. We discuss symmetries in more detail below, here note only that below $\beta_{c,1}$ the trivial constant encoder is fully insensitive to re-labelling, or permuting, of the variables $h$, and without loss of generality can be written as $P(h|v) = P(h|r_v) = 1/|H|$. In order to learn \emph{any} information whatsoever, this symmetry has to be broken. At $t$ small enough, $\langle r_v \rangle_h$ tends to zero, and we can therefore expand: 
\begin{align}
\label{eq:reducedIB_around_betac1}
P(h|r_v) &= \frac{1}{|H|}+t b_{r_v}(h) \\ \nonumber
\sum_h b_{r_v}(h) &= 0, 
\end{align}
where the second equation ensures proper normalization of the conditional probability. 
Plugging the above into the left-hand side of Eq.\ref{Eq:reducedIB} and expanding the right-hand side in $t$ we get, to lowest order: 
\begin{align}
|H|^{-1} + t b_{r_v}(h) &= |H|^{-1} + \beta \epsilon^2  t r_v \sum_{v'} P(v') r_{v'} b_{r_{v'}}(h) 
\end{align}
Examining the above one finds that a $b(h)_{r_v}$ which is constant in $r_v$  is always a solution, for any $\beta$. This simply implies that any $P(h|r_v)=P(h)$ is a solution to the IB equation for all $\beta$. Moreover, such solutions are globally optimal before the first phase transition. This is a special case of a more general phenomenon: when some symbols $h$ share exactly the same dependence on V, there is a manifold of equivalent solutions reflecting the freedom to re-distribute probabilities between these symbols $h$, in particular to assume them maximally symmetric.

In order to construct an explicit closed-form solution and to later verify it numerically, let us proceed by focusing on $|H|=2$. Thus the compressed variable has two states and can be thought of as a single spin degree of freedom: $h=\pm 1$. 
The above equations now become:
\begin{align}\label{eq:reducedIB_around_betac2}
b_{r_v}(h) &= \beta \epsilon^2 r_v \sum_{v'} P(v') r_{v'} b_{r_{v'}}(h) \\ \nonumber 
\sum_h b_{r_v}(h) &= 0 .
\end{align}
The first equation, with a fixed $h$, when viewed as linear equation on the vector space spanned by the values of $r_v$ and equipped with an inner product weighted by $P(v)$, has two solutions: the aforementioned constant-$r_v$ vector is a solution for any $\beta$. The vector $b_{r_v}=r_v$ is a solution for $\beta \epsilon^2 = 1$. This can be arranged into a solution for both $h$ by taking $b_{r_v}(h) = r_v h$. The $\beta$ at which this soft perturbation to the uniform encoder appears, marks the first critical $\beta$:  
\begin{align}\label{eq:betac1}
\beta^{-1}_{c,1} &= \epsilon^2 + o(\epsilon^{2})
\end{align}
where through $o(\epsilon^2)$ we re-introduced possible corrections coming from $(\lambda_{n>1}/\lambda_0)^{L_B}$, all scaling as higher powers of $\epsilon$. These corrections exhibit a faster exponential decay as a function of $L_B/L$ and are hence negligible for $L_B/L \gg 1$. Keeping this ratio fixed and taking $L \rightarrow \infty$, one can use the fact that:
\begin{align}
\lambda_1/\lambda_0 \xrightarrow[]{L \rightarrow \infty} e^{-2\pi \Delta_1/L},
\end{align} 
where $\Delta_1$ is the CFT scaling dimension associated with the leading primary operator. We stress though, that all the results of this section apply to any $L$ (provided $L_B \gg L$) and do not rely on having a CFT, apart from the association between the transfer matrix eigenvalues $\lambda_n$ and the scaling dimensions $\Delta_n$ in the large $L$ limit, which is important for interpretation.

To obey normalization of $P(h|v)$, this $r_v$-linear vector has to be added with opposite signs to form $P(h=\pm 1|r_v)$. 
Examining Eq.\ref{Eq:reducedIB} together with the assumption that $P(h)$ is constant leads to the following solution ansatz:   
\begin{align}\label{eq:encoder1}
P(h=\pm 1 | r_v) &= \frac{e^{h m(t)r_v}}{2 \cosh(m(t)r_v)},
\end{align}
where $m(t<0)=0$ and $m(t>0)>0$. To determine $m(t)$ we plug the above encoder into Eq.\ref{Eq:reducedIB}:
\begin{align}
\frac{e^{h m(t)r_v}}{2 \cosh(m(t)r_v)} &= \frac{1}{2} \frac{e^{\beta \epsilon^2  r_v \langle r_v \rangle_h}}{ \cosh(\beta \epsilon^2 r_v |\langle r_v\rangle_h|)}
\end{align}
where we have used $P(h)=const.$ and the fact that $\langle r_v \rangle_h = h |\langle r_v \rangle_h|$. Clearly, if both numerators are equal, than the denominators would agree as well. Thus, we compare the logarithm of both numerators and obtain: 
\begin{align}\label{eq:encoder2}
hm(t) r_v = \beta \epsilon^2 r_v \langle r_v \rangle_h.
\end{align}
The average on the right-hand side is evaluated with $P(r_v|h)=P(r_v)P(h|r_v)/P(h)$ and yields $\langle r_v \rangle_h = h |\langle r_v \rangle_{h=1}|$ and therefore:
\begin{align}\label{eq:encoder3}
m(t) = \beta \epsilon^2 |\langle r_v \rangle_h|
\end{align}
An expansion of the right-hand side in $m(t)$ up to third order yields:
\begin{widetext}
	\begin{align}\label{eq:mt1}
	m(t) &= \beta \epsilon^2 \sum_{r'_v} \frac{ P(r'_v) r'_v \left[1+m(t)r'_v + \frac{1}{2}m^2(t) {r'_v}^2 + \frac{1}{6} m^3(t) {r'_v}^3 \right]} {1+\frac{1}{2} m^2(t) {r'}^2_v} \\ \nonumber & = \beta\epsilon^2 \sum_{r'_v} P(r'_v) r'_v \left[1+m(t)r'_v +\frac{1}{2} m^2(t) {r'}^2_v + \frac{1}{6} m^3(t) {r'}_v^3\right] \left[1 - \frac{1}{2} m^2(t) {r'}^2_v\right] \\ \nonumber 
	&= \beta\epsilon^2 \left[m(t)\langle r^2_v \rangle +\frac{1}{6}m^3(t) \left(\langle r_v^4\rangle -3\langle r_v^4\rangle\right)\right] + O(m^4)
	\end{align}
\end{widetext}
We thus finally obtain: 
\begin{align}\nonumber
(\beta-\beta_c) \epsilon^2 &- \frac{\langle r^4 \rangle}{3}m(t)^2 = 0 \\ \label{eq:betac2}
m(t) &= \sqrt{\frac{3 (\beta-\beta_c)}{\langle r^4 \rangle \beta_c}}
\end{align}

We have thus an explicit analytical (and closed-form) solution for the encoder and the critical $\beta_{c,1}$ in terms of CFT quantities which can be computed in the transfer matrix formalism, in terms of TM eigenvalues and eigenvectors (either analytically, or, as is common, by numerical TM diagonalisation). 

We compare this theoretical prediction for the behaviour of the IB solutions to the ones obtained numerically (\emph{i.e.}~by feeding Monte Carlo samples of the system to the IB solver as the input probability distribution, as we would do with any other data, physical or not, in a generic compression problem). As seen in Fig.\ref{fig:fig3} in the main text, when presented with the data for the 2D critical Ising model on a cylinder, the analytical solution is in excellent agreement with numerics. The IB encoder $P(h|v)$ does indeed depend on $\partial V$ only, and it assigns the value of the coarse-grained spin $h$ based on the magnetization of the configuration of spins in $\partial V$, that is it depends on the physically most relevant operator, in exactly the predicted fashion.

\section{IB and the physical symmetries}
Here we derive several results regarding IB in the presence of physical (or model) symmetries in the data, some of which were mentioned in the main text. Along the way we also briefly review the necessary results on the internal (or structural) symmetries in IB. In this, and in the formal tools we use we follow Refs.\cite{parkerentropy,parkernips}. 

\begin{figure*}[tp]
	\centering
	\vspace{-1.5cm}
	\includegraphics[width=0.75\textwidth,trim={0 2cm 0 0}]{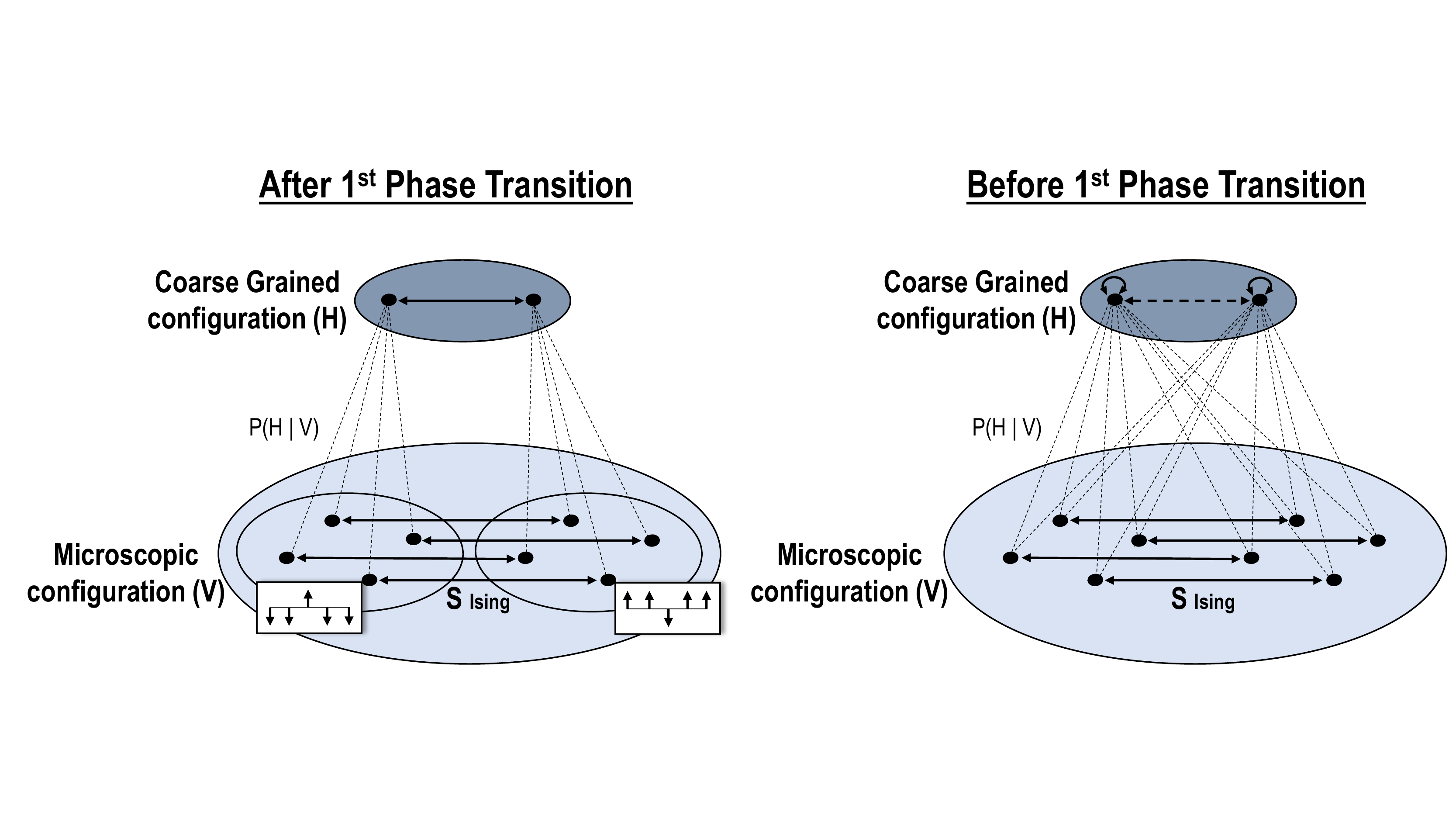}
	\caption{\textbf{IB and the physical symmetries}: the physical system, \emph{e.g}~an Ising model, is invariant under a symmetry. The orbits of the group action are depicted with arrows connecting the symmetry related configurations in $V$ (here: by global $Z_2$ spin-flip). \textbf{Right:} Before the first IB transition the trivial encoder maps every $v \in V$ equally likely to both $h \in H$. This is consistent with a trivial action of the $Z_2$ symmetry on $H$. All symbols $h$ are thus equivalent and connected by the action of the structural permutation symmetry (dashed line in $H$). \textbf{Left:} After the transition, distinct elements $v$ would be preferentially mapped to particular symbols $h$. Elements $v$ which are related by the physical (here: Ising) symmetry are mapped to $h$ in a manner which generates a non-trivial action of the symmetry on $H$. Since the symbols $h\in H$ become inequivalent, the structural IB symmetry is broken.}
	\label{fig:ibrsmi}
\end{figure*}

Phase transitions in IB refer to values of $\beta$ where some non-analyticity appears in ${\rm L}_{IB}$ as a function of $\beta$. These transitions often come from breaking of IB structural/intrinsic symmetries although, in principle, other transitions (\emph{e.g}~saddle-nodes \cite{parkerentropy}) are possible where the structural symmetries remain the same. The IB structural symmetries consist of permuting the classes, or elements, of $H$ as well as re-weighing the conditional probabilities: taking $P(h|v) \rightarrow P(h|v)(1+\alpha_{h})$ while maintaining the normalization constraint $\sum_h P(h|v)(1+\alpha_h)=1$ for all $v$ and the value of relevant information preserved. This large continuous freedom requires some form of ``gauge" fixing. We adopt the natural prescription of \cite{parkerentropy,parkernips}, and work with encoders $P(h|v)$ which are as uniform as possible in $H$. Namely, we always re-weigh them so that for any symbols $h,h'$ for which $P(h|v) = c P(h'|v) 
\,\ \forall v$ with $c \in \mathcal {R}_+$, the conditional probabilities are shifted to be identical. For example, this means that for $\beta < \beta_{c,1}$ the encoder is completely symmetric: $P(h|v) = 1/|H|$. 

The above gauge freedom can be understood with a simple example: consider compressing information into a code with exactly three symbols $h_{1,2,3}$, with the code assigning them based on the inputs $v$. Imagine a code which \emph{never} assigns the symbol $h_3$ to any input, and another one, in which, given an input $v$ which should be mapped to $h_2$ by the previous code we instead randomly assign $h_2$ or $h_3$ with probability $p_2+p_3=1$. Exactly the same information can be retrieved from both of these codes, for any $p_2$, which represents the ``gauge freedom". Given the possibility of using $|H|$ symbols, we use all of them, but only a few are used nontrivially (that is $P(h|v)$ actually depends on $v$), the rest appear entirely randomly and independent of the inputs, with equal probability, and thus carry no information whatsoever. They can be thought of as being ``unresolved", as their probability does not depend on any feature of the data. The advantage of this formulation over simply removing unresolved symbols from the formalism, is that in IB phase transitions new symbols become ``resolved", that is the encoder starts using them nontrivially to track some additional feature of the data, and in this process their conditional probability distribution acquires dependence on $v$, breaking the symmetry of permuting all unresolved symbols.

Given the above re-weighing choice, IB transitions, unless fine-tuned, appear as they do in physics via first or second order symmetry-breaking transitions, where it is the permutation symmetry that is being broken at the point of transition. In IB terminology these are called subcritical and supercritical pitchfork bifurcations \cite{parkernips}, respectively. Provided $|H|$ is taken to be $2|V|-1$ or more, first order transitions are also excluded \cite{etam}.

Despite the efforts to classify structural transitions, the question of how well \emph{physical}/model symmetries present in the data are reflected or preserved in the IB transitions has, to the best of our knowledge, not been explained. Given fundamental role of symmetries in physics, this is an important point. Below we detail two contributions we make towards clarifying this issue. 

Assuming that \textbf{(a)} $\beta < \beta_{c,2}$, \textbf{(b)} $|H|$ is large enough such that taking $|H| \rightarrow |H|+1$ does not lead to a better IB solution (we are not constrained by a small code alphabet), \textbf{(c)} the first IB transition is second order (implied by \cite{etam}), and \textbf{(d)} this transition cannot be split into two separate transitions using a perturbation to $P(v,e)$ respecting the physical symmetry (\emph{i.e.}~no fine-tuning),  we show that the following holds true: 
\begin{align}
\label{AppEq:sym_rel}
P(e,v) = P(se,sv) &\Rightarrow P_\beta(h|v) = P_\beta(\phi_s h | s v)
\end{align}
Here the element $s$ of the symmetry group $\mathcal{S}$ acts by permutation on the configurations of the system (with the action denoted by multiplication), and invariance under this symmetry is expressed by equality of their probabilities; $\phi_s$ is a subgroup of the permutation group on $H$ which obeys $\phi_s \phi_{s'}=\phi_{ss'}$ (\emph{i.e.} it is a permutation representation of the symmetry $\mathcal{S}$, possibly a trivial one). Eq.\ref{AppEq:sym_rel} states that under the assumptions stipulated above, the optimal IB encoder carries a representation of the physical symmetry (thus ensuring that the coarse-grained probability $P(h)$ also does).
Fig.\ref{fig:ibrsmi} shows a schematic picture of the symmetry action on $V$ and $H$, before and after the first IB transtion.

We also consider the case of small $|H|$, such that IB is constrained from finding an optimal solution. We construct an example with a $Z_4$ physical symmetry and $|H|=2$, where Eq.\ref{AppEq:sym_rel} is violated, implying that that encoder breaks the physical symmetry.

\subsection{IB and symmetries near $\beta_{c,1}$} 
For $\beta<\beta_{c,1}$, following our choice of gauge, $P(h|v)=1/|H|$. Thus $P(h|v)=P(h|sv)$ and hence Eq. (\ref{AppEq:sym_rel}) holds with $r_s = id$ being the trivial representation. 

To study $\beta>\beta_{c,1}$ we follow the approach of Refs.\cite{parkerentropy,parkernips}. The main idea is to study the stability of the minima of the IB Lagrangian $\mathcal{L}_{IB}$, which correspond to the solutions for the optimal encoder $P_\beta(h|v)$. To this end the functional $\mathcal{L}(P,\lambda,\beta) = \mathcal{L}_{IB}(P,\beta) + \sum\nolimits_v \lambda_v (\sum\nolimits_h P(h|v)-1)$ is introduced, with $\lambda_v$ the Lagrange multipliers for the constraint enforcing normalization of the conditional probabilities $P$. The stable local solutions are such for which all of the eigenvalues of the Hessian $\Delta_{P,\lambda}\mathcal{L}(P,\lambda,\beta)$ have negative real parts. Here $P$ is a vector of conditional probabilities $P(h|v)$ of dimension $|H|\cdot|V|$, and $\lambda$ is a vector of $\lambda_v$ of dimension $|V|$; we differentiate (twice) with respect to all of the components.

The strategy then is to consider the Hessian at $\beta = \beta_{c,1}$, analyze its kernel and how the physical symmetry manifests itself on the space spanned by the eigenvectors corresponding to eigenvalue(s) crossing from positive to negative at $\beta=\beta_{c,1}$ (the crossing eigenvalues). Following \cite{parkerentropy}, we split these crossing eigenvalues into groups belonging to each element of $H$ and show that these smaller groups generate irreducible representations of the physical symmetry, provided the transition is not fine-tuned. We then explicitly construct a globally optimal encoder at $\beta \gtrapprox \beta_{c,1}$ which obeys \ref{AppEq:sym_rel}. Having shown that the symmetry remains unbroken just after $\beta_{c,1}$, and given that $\mathcal{L}_{IB}$ obeys that symmetry, by continuity we establish our claim for all $\beta < \beta_{c,2}$.

\begin{figure}[tp]
	\centering
	\includegraphics[width=6cm, height=5cm]{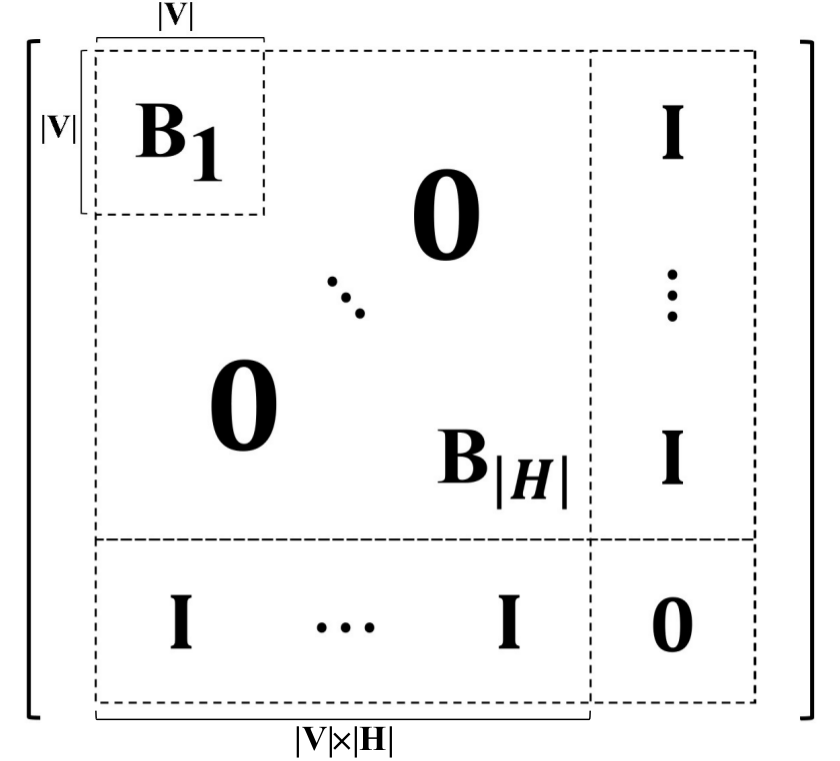}
	\caption{The structure of the Hessian $\Delta_{p,\lambda}\mathcal{L}(p,\lambda,\beta)$}
	\label{fig:hessian}
\end{figure}

We first note that the IB Lagrangian has the property that it splits into a sum of different contributions from distinct $h$, namely:  
\begin{align}\nonumber
\sum_h &\left[ \sum_v P(h|v)p(v)\log(P(h|v)) -  (1-\beta)P(h)\log(P(h)) \right. \\ \nonumber
- \beta&\left. \sum_{e,v} P(h|v)P(v|e)P(e)\log(\sum_v P(h|v)P(v|e)) \right] \equiv\\ 
\equiv &\sum_h L_h
\label{AppEq:Lh}
\end{align}
where $L_h$, implicitly defined above, contains only a single $h$. It is also symmetric under $P(h|v) \rightarrow P(h|sv)$, assuming the l.h.s. of Eq. \ref{AppEq:sym_rel}. The interdependence between $P(h|v)$ with different $h$ enters solely through the normalization requirement $\sum_h P(h|v)=1$, which can be maintained by adding a Lagrange multiplier term $\sum_{h} \sum_v \lambda_v [P(h|v)-1]$. The Hessian with respect to (w.r.t.) $P(h|v)$ and $\lambda_v$ has a special structure \cite{parkerentropy}. Below $\beta_{c,1}$ it consists of $|H|$ equal block matrices $B$ of size $|V|$ on the diagonal, along with identity matrices of size $|V|$ on the last column and row, related to the Lagrange multiplier ensuring the normalization (see Fig.\ref{fig:hessian}). The matrix $B$ is simply the Hessian of $L_h$ w.r.t. $P(h|v)$ around the trivial encoder (which is independent of $h$). Generically, for higher $\beta$, only the blocks $B_h$ corresponding to unresolved symbols $h$ are identical, their equality being the consequence of the permutation symmetry of these symbols.

An IB phase transition where new stable solutions appear, and which is second order, necessitates a non-empty set of eigenvalues of the Hessian (the aforementioned crossing eigenvalues) changing from positive to negative -- at the transition the kernel of the Hessian changes. We now argue that their existence also implies the existence of a smaller set of crossing eigenvalues \emph{within each block} $B$. Let us consider an eigenvector $w$ of the Hessian, corresponding to a small eigenvalue $\epsilon$ of order $\mathcal{O}(\beta_{c,1}-\beta)$ (possibly one of many such eigenvectors). Utilising the knowledge of the structure of the full Hessian (see Fig.\ref{fig:hessian}) we write it in block from as $w = [w(h_1),\ldots,w(h_{|H|}),\eta ]^{\mathrm{T}}$. The eigenvalue equation corresponds to a set of $|H|+1$ vector equations: 
\begin{align}\label{eq:hessian1}
B w(h) + \eta &= \epsilon w(h) \\ \nonumber
\sum_h w(h) &= \epsilon \eta 
\end{align}
Summing the first line over all $h$ we find $B \eta = (-|H| \epsilon^{-1}+\epsilon) \eta$, which at small enough $|\epsilon|$, and for a bounded $B$, has only the solution $\eta=0$. Consequently, we find that for any $h$ we have $B w(h)= \epsilon w(h)$, \emph{i.e.}~the subblocks of crossing eigenvectors of the full Hessian define eigenvectors for the individual blocks, which have vanishing eigenvalues (\emph{i.e.}~which belong to their kernels). In general there may exist multiple distinct crossing eigenvectors of the full Hessian, and consequently of the blocks. We denote them by $w_i$.  

To proceed, note that in the IB formalism the cardinality of the alphabet $H$ is not restricted. We therefore wish to factor out the dependence of the IB Lagrangian on $|H|$. Close to the first phase transition the encoder can be written as a perturbation of the trivial (maximally symmetric) one \cite{parkerentropy}:
\begin{align}\label{eq:perturbedencoder}
P(h|v) &= |H|^{-1}\left[1 + |H| \sum\nolimits_i c_i(h) w_{i,v}\right] \\ \nonumber &= |H|^{-1}\left[1 + \sum\nolimits_i \tilde{c}_i(h) w_{i,v}\right]
\end{align}
where $w_{i,v}$, viewed as vectors in the $v$ indices for fixed $h$, are the crossing eigenvectors. Plugging the above r.h.s. into Eq.\ref{AppEq:Lh}, and using $\log(|H|^{-1}\left[...\right])=\log(|H|^{-1}) + \log(\left[...\right])$, one finds that all the terms proportional to $\log(|H|^{-1})$ in $L_h$, vanish. In the remaining terms $|H|^{-1}$ enters explicitly, but only as an overall scaling factor. Note that Eq.C1 is a special case of the above. With this simplification, near the first transition $|H|L_h$ can be written as:
\begin{align}
\tilde{L}_h &\equiv |H| L_h = -\epsilon A \sum_{i} \tilde{c}_i(h) \tilde{c}_i(h) + \Delta \tilde{L} [\tilde{c}_1(h),...,\tilde{c}_N(h)]
\end{align}
where $\Delta \tilde{L}$ is cubic or above in $\tilde{c}_i(h)$ (consistent with second order transitions).

We proceed by analyzing the potential minima of all the terms $\tilde{L}_h$. At $\epsilon<0$, all $\tilde{L}_h$ have a single minimum at $\tilde{c}_i(h)=0,\tilde{L}_h=0$ and thus the encoder is uniform and trivial by Eq.\ref{eq:perturbedencoder}. For a second order permutation-symmetry-breaking transitions (supercritical pitchfork transition, in the terminology of Ref.\cite{parkerentropy}), at $\epsilon \gtrapprox 0$ the rescaled Lagrangian $\tilde{L}_h$ develops $N>0$ global minima (H-minima) of $\tilde{L}_h$ where $\tilde{L}_h = \tilde{L}_{min} < 0$, which will later be used to construct the global minima of $\mathcal{L}_{IB}$.  Each of these H-minima is defined by some $\tilde{c}_i(h) \neq 0$. We label the coefficients corresponding to these distinct minima by $\tilde{c}^{0,n}_i(h)$ with $n \in \{1,\ldots,N\}$, which implicitly depends on $\epsilon$. Note that around $\beta_{c,1}$ the set of H-minima (or equivalently the coefficients) are the same for different $h$ and $\tilde{L}_h$, because we perturb around the fully symmetric solution.

Next we claim that any encoder which uses only the H-minima, namely one defined as $P(h|v)=|H|^{-1}[1 + \sum\nolimits_i \tilde{c}^{0,n_h}_i(h) w_{i,v}]$, which is properly normalized,
is globally optimal among all choices of coefficients $\tilde{c}_i(h)$ as well as sizes of $|H|$ which yield a normalized probability distribution. Indeed, $\min(\mathcal{L}_{IB}) = |H|^{-1} \sum_{h\in H} \tilde{L}_{min}=\tilde{L}_{min}$. Next due to $\tilde{L}_{min}$ being $|H|$ independent, one has that $\tilde{L}_{min} \le |H'|^{-1}\sum_{h\in H'} \tilde{L}_h$, since $\tilde{L}_h \ge \tilde{L}_{min}$.

We have thus shown that at the transitions, a set of H-minima appear in each $\tilde{L}_h$ which, if composed in a way that obeys normalization, result in a globally optimal encoder. Next we discuss how the symmetry $\mathcal{S}$ acts on the H-minima. This will be used to construct a normalized \emph{and symmetric} encoder which uses only the H-minima. Given an underlying symmetry $\mathcal{S}$, whose elements $s$ act as permutations of elements $v$ of $V$, one finds that the blocks $B$ obey the symmetry via $B_{v,v'}=B_{sv,sv'}$. Hence the crossing eigenvectors within each $H$ block $w_{i,v}$ viewed as vectors in the space spanned by elements $v$, transform as some real representation of the symmetry, namely $w_{i,s v}=\sum_{j} s_{ij}w_{j,v}$ (with $\sum_j s_{ij}s_{jk}=\delta_{ik}$). It can then easily be shown that $\tilde{c}_i(h)$ transforms as $s \cdot \tilde{c}_i(h) = \sum_j s_{ji} c_j(h)$ and that $s \in \mathcal{S}$ acting on the $n$-th H-minimum $\tilde{c}^{0,n}_i(h)$ leads to an H-minimum, say $\tilde{c}^{0,m}_i(h)$, and so we write that $s(n)=m$. Consequently, we have a group action of the symmetry on the set of different H-minima. We note in passing that this action will later determine $\phi_s$.

Let us next assume that $s_{ij}$ for all $s\in \mathcal{S}$ form an irreducible representation on the crossing eigenvectors (i.e. those with eigenvalue $\epsilon$). Notably, this is also the generic case, as one does not expect to find degenerate sets of eigenvalues beyond what is implied by symmetry. If the latter does happen it implies the transition can be split into two nearby transition using a symmetry respecting perturbation to $P(v,e)$.

Consider first the case where the representation $s_{ij}$ is trivial. Here much of the machinery we developed is not needed since any encoder just after the transition, in particular the globally optimal one, would depend on $v$ via $w_{i,v}$, and since the latter is invariant under the symmetry, Eq.\ref{AppEq:sym_rel} is obeyed with a trivial $\phi_s$ equal to the identity permutation ($id$). 

We thus turn to the case of a non-trivial irreducible representation. Here we take $|H|=N$, associate each $h$ with a specific $n_h$, and henceforth drop the distinction between $n$ and $h$, writing $\tilde{c}^0_i(h)$ as shorthand for $\tilde{c}^{0,n_h}_i(h)$. We further split the action of $\mathcal{S}$ on the H-minima into orbits $O$, each orbit understood as a set of $h$ values corresponding to a set of H-minima. We claim that within each orbit $\sum_{h \in O} \tilde{c}^0_i(h) = 0$ for all $i$. Indeed, due to the orbit being closed under any $s \in \mathcal{S}$ we have :
\begin{align}
\sum_{h \in O,i} \tilde{c}^0_i(h) w_{i,v}&=\sum_{h \in O,i} \tilde{c}^0_i(s(h)) w_{i,v} \\ \nonumber 
&=\sum_{h \in O,i} \tilde{c}^0_i(h) w_{i,sv} 
\end{align}
thus if the l.h.s. is not zero, the r.h.s. implies we have found a vector $w_i$ in the set of crossing eigenvalues, which is invariant under the action of any $s$. This would lead to a contradiction, as the representation was assumed to be irreducible and nontrivial.

Finally, we write our globally optimal, normalized, and symmetric encoder just after $\beta_{c,1}$:
\begin{align}
P(h|v) &= |H|^{-1}[1 + \sum_i \tilde{c}^0_i(h) w_{i,v}]
\end{align}
where the dependence on $\beta$ enters implicitly via $\tilde{c}^0_i(h)$. It can be verified that it obeys Eq. (\ref{AppEq:sym_rel}) with $\phi_s$ being the action of $\mathcal{S}$ on the H-minima: $s(h)=h'$. Furthermore, it is normalized since: 
\begin{align}
\sum_h P(h|v) &= 1 + |H|^{-1}\sum_{O}\sum_{h \in O} \sum_i \tilde{c}^0_i(h) w_{i,v} = 1
\end{align}
Lastly, as this encoder uses only H-minima it is globally optimal. 

Had we taken $|H|<N$ such a solution would not be possible, and in such circumstance IB can potentially break physical symmetries. We provide such an example below (though at high $\beta$). Note though, that it is always possible to increase $|H|$ until it no longer improves $\mathcal{L}_{IB}$, thereby avoiding these constrained settings. 

We conjecture that IB, unless constrained or tuned in an adversarial fashion, respects physical/model symmetries for all values of $\beta$. In particular we have never encountered a numerical example where this does not hold for large enough $|H|$. Notably, for unrestricted $|H|$ there is no obvious competition between learning an optimal encoder and maintaining the symmetry which could encourage such physical/model symmetry breaking.

\subsection{Potential breaking of the physical symmetry in constrained IB at large $\beta$}

The results of the previous section imply that for large enough $H$ the encoder would carry a representation of the physical symmetry.
Given the fact that in practical implementations the size of alphabet $H$ would often be fixed, and the symmetry group may possibly not be fully known, it is interesting to ask what happens in the case $|H| < N$. 

The question is whether, given possible sizes of permutation representations of the symmetry and some fixed $|H| < N$, more information is \emph{necessarily} retained when the encoder generates the action of one of those representation on $H$, or not. To make intuitive why breaking the symmetry could be favorable, consider the following example. Let $n_1 > |H| > 1$ be the size of the smallest nontrivial permutation representation of the physical symmetry, \emph{i.e.}~we are given  more symbols than needed to ``fit" the action of the trivial representation on $H$ (which maps everything to one symbol), and not enough to fit a nontrivial one. It seems natural that not using the available $H$ symbols in the encoder is wasteful.

Here we provide an explicit example where Eq.\ref{AppEq:sym_rel} is violated at $|H|<N$ in the limit $\beta \rightarrow \infty$. Such examples are easier to construct if $|H|$ is incompatible with the dimension of any irreducible representation of $\mathcal{S}$, as mentioned above. Here, however, we give a less trivial example with a $Z_4$ symmetry and $|H|=2$, which intuitively at least could fit a two-dimensional representation of the symmetry. 

For $\beta \rightarrow \infty$, the IB Lagrangian simplifies to maximizing $I(H;E)$. Intuitively, in this limit the solution should always be a deterministic encoder (\emph{i.e.}~one for which $P(h|v) \in \{0,1\}$) and defines a bona fide function $f:v \rightarrow h$. This in fact follows from convex optimization arguments, as shown in \cite{Hassanpour2017}. Next we write the mutual information as difference of entropies: 
\begin{align}\label{app_eq_ent_diff}
I(H;E)=S(E) - S(E|H),
\end{align}
where: 
\begin{align}
S(E|H) &= \sum\nolimits_h P(h) S(E|h) \\
S(E|h) &=-\sum\nolimits_e P(e|h)\log(P(e|h)).
\end{align}
Note that due to the deterministic nature of $P(h|v)$:
\begin{align}\nonumber
P(v|h)=P(h|v)P(v)/P(h)=P(h)^{-1} P(v) \delta_{f(v),h}.\end{align} 
Therefore: 
\begin{align}
P(e|h) &= \sum_{v}P(e|v)P(v|h) \\ \nonumber  &=\sum_{v|f(v)=h}\frac{P(v)}{P(h)}P(e|v)\equiv P(e|V(h)),
\end{align} 
where by $V(h)$ we denoted the pre-image of $h$ under mapping $f$.
By the above equations we thus seek to group the elements $v$ into $h$-clusters, such that the entropy of $E$ averaged over the different clusters is minimized. Formally, we minimize:
\begin{align}\nonumber
\sum_h P(h)& S(E|h)= \\ \nonumber
 &=-\sum_h P(h) \sum_e P(e|V(h) \log(P(e|V(h))) \\  &\equiv \sum_h S(E|V(h))P(h) \label{eq:app_sevh}
\end{align}
Let now $V,E=\{0,1,2,3\}, H=\{0,1\}$, $S=Z_4$ and let
\begin{align}\nonumber
P(e|v)=P(v|e)= \frac{1}{3}[\delta_{e,v} + \delta_{e,(v+1) \% 4} + \delta_{e,(v+2) \% 4}],\end{align} where $\%$ denotes the modulo operation, and let $P(v)=P(e)=1/4$ for all $e$,$v$. We want to find the two disjoint sets $V_1$ and $V_2$ (which are mapped to distinct $h$) which minimize $S(E|V_1)P(h=0)+S(E|V_2)P(h=1)$, \emph{i.e.}~equation \ref{eq:app_sevh}. Up to an action of $Z_4$ there are only two distinct partitions of $V$ which are maintained by the action of the only nontrivial subgroup of $Z_4$, i.e. $Z_2$, and thus which could be compatible with an encoder producing the action of $Z_2$ on $H$: $V_1 = \{0,2\},V_2 = \{1,3\}$ and $V_1 = \{0,1\},V_2=\{2,3\}$. For these two partitions of $V$ we have that: 
\begin{align}\nonumber
S(E|V_1)P(h=0)&+S(E|V_2)P(h=1)= \\ \nonumber
&= \frac{1}{2}[S(E|V_1)+S(E|V_2)].
\end{align} 
We compare these to the following non-symmetric choice $V_1 = \{0\}, V_2=\{1,2,3\}$. All together these exhaust all symmetry-distinct choices of the sets. For the first symmetric choice, one has $P(e|V_{1})$ being equal to $e$ drawn uniformly from $\{0,0,1,2,2,3\}$ leading to $S(E|V_{1/2})=1.3296$. For $V_1$ of the second symmetric choice, we get $e$ drawn from $\{0,1,1,2,2,3\}$ leading to the same value of $S(E|V_{1/2})$. For $V_1$ of the non-symmetric choice, we get $e$ drawn uniformly from $\{0,1,2\}$ leading to an entropy $S(E|\{0\})=1.0986$. Last for $V_2$ of the non-symmetric choice, we get $e$ drawn uniformly from $\{0,0,1,1,2,2,3,3,3\}$ leading to $S(E|\{1,2,3\})=1.3689$. As $1.3296 > 0.25\cdot1.0986+0.75\cdot1.3689=1.3013$ we find that the non-symmetric choice is optimal. 

While the $\beta \rightarrow \infty$ example can be evaluated on the back of an envelope, we verified numerically that for this example distribution the symmetry breaking holds for $\beta$ all the way down to the first IB phase transition. Interestingly, taking a larger alphabet $|H|>2$ improves the IB Lagrangian value $\mathcal{L}_{IB}$ of the best encoder at $\beta \geq \beta_{c,1}$, until at $|H|\geq 4$, \emph{i.e.}~at $|H|\geq N$ it reaches its optimal value and remains unchanged by further increasing $|H|$. 

The lessons we take from the above example are that \textbf{(a)} for $H$ of insufficient size, smaller than the size of the (relevant) symmetry group, the symmetry can be broken by the encoder and \textbf{(b)} in order to ensure this does not happen one should choose the minimal $|H|$ at which, for a fixed $\beta \geq \beta_{c,1}$, the value of the IB Lagrangian $\mathcal{L}_{IB}(P_\beta(h|v))$ reaches its optimal value. Recall also that $\beta_{c,1}$ can be obtained in a model-agnostic way by studying the kernel of the $B$ block of the Hessian. 

\section{Extracting symmetries from a numerically obtained encoder}
Here we discuss the possibility of using the symmetry-maintaining properties of the optimal encoder, \emph{viz.}~Eq.\ref{AppEq:sym_rel} to extract the physically relevant symmetries from the data. 

Let us examine the situation where $P(v,e)$ possesses an unknown symmetry $s \in \mathcal{S}$, such that $P(sv,se)=P(v,e)$, and an unknown action $\phi_s$ on the $h$ variables. The symmetry may be unknown since it involves a complicated combination of microscopic degrees of freedom, or because $\mathcal{S}$ is part of a much larger symmetry group from which we wish to sift out the most relevant subgroup. In addition, even if $\mathcal{S}$ is known, its action on $h$ would depend on how it combines the relevant variables and may potentially need to be extracted numerically. We discuss how both $\mathcal{S}$, $\phi_s$, and their actions on $v$ and $h$, respectively, can in principle be identified. 

As an instructive example, consider the Ising model used in the main text. We deliberately split, however, each Ising spin $\sigma_i$ into to product of two auxiliary spins $\sigma_i = \tau_{i1}\tau_{i2}$. The energy of the system remains the same (in terms of the original spins).
By construction, this model has a huge amount of spurious symmetry: an extra $Z_2$ symmetry per each site $i$, given by $I_{i,1}I_{i,2}$ (where $I_{i,*}$ is a spin-flip operator for the variable $\tau_{i*}$). This symmetry does not flip the Ising spin $\sigma_i$ and so bears no influence on the long-range properties of the system, -- the physical Ising symmetry is artificially obscured in the model phrased in $\tau$ microscopic variables.

Assume now we are given the solution to the IB problem $P_{\beta}(h|\tau)$ with $h=\pm 1$ for $\beta > \beta_{c,1}$, which depends on the Ising magnetization on the boundary $\sum_{i \in \partial V_R} \tau_{i1}\tau_{i2}$ (as per arguments in the main text and in the appendices above). Since for this encoder: 
\begin{align}
P_\beta(h|\tau)=P_\beta(h|I_{i,1}I_{i,2} \cdot\tau)
\end{align}
it generates a \emph{trivial} representation of all the extra $Z_2$ symmetries. Since it couples to the physical magnetization, however, it generates a \emph{faithful} $Z_2$ representation of the ``hidden" Ising symmetry (which we can choose as $s=\Pi_{i \in V} I_{i1}$), namely $P(h|\tau)=P(-h|s\tau)$. We would like to provide a prescription for identifying the relevant symmetry $\mathcal{S}$ and its action on $h$. 

The first step is to obtain an estimate of the $P_{\beta}(h) = \sum_v P_{\beta}(h|v)P(v)$, by numerically sampling $v$. Using the results on the general structure of the optimal encoder, it suffices to take $v \in \partial V_R$. The symmetry of the encoder $P(h|v)=P(\phi_s h | s v)$ implies the symmetry in the coarse-grained variables: $P_{\beta}(h)=P_{\beta}(\phi_sh)$. This allows to group the equiprobable elements $h$ into sets, whose elements are potentially related by an action of $\phi_s$ for some $s \in S$. These sets, forming a partition of $H$, are putative orbits of $\phi_s$.
In the above example the set simply contains both $h=\pm 1$. 

Consider now configurations $v_+ \in \partial V_R$ for which $P(+1|v_+)$ is non-zero (generally this yields all configurations) and similarly so for $v_-$. Next reconstruct the action of the symmetry element $s$ on $V$ by demanding that: \textbf{1.} $s$ maps the set of all $v_+$ to the set of all $v_-$, \textbf{2.} $P(+1|v_+)=P(-1|s v_+)$, \textbf{3.} $s$ applies the same onsite permutation across all sites (\emph{i.e.}~$s$ is spatially homogeneous). Notably, committing to such homogeneous $s$ is allowed provided we focus on global symmetries. Conveniently, it also makes the space of potential permutations much smaller, in the sense that it is independent of the number of sites. Following this $s$ can be found using a brute force scan. 

In our example, provided $\beta$ is finite, the set of all $v_+$ is simply $\partial V_R$ and $s_1=\Pi_{i \in \partial V_R} I_{i1},s_2=\Pi_{i \in \partial V_R} I_{i2}$. It is then easy to infer from the symmetry requirement $P(h|x)=P(\phi_s h|sx)$ that $s_1$ (and $s_2$) have a $Z_2$ action on $H$ whereas $s_1 s_2$ has a trivial action. We have thus exemplified how to identify qualitative information from the numerically obtained encoder: the size of the representation of the relevant symmetry was deduced from the number of equally probable symbols $h$, and its action on $v$ was given by the above permutations $s$ obeying the symmetry constraints. Using the action on $v$, $\phi_s$ acting on $H$ can also be deduced.  

Had we considered a larger set of symmetric $h\in H$'s (say $|H|=3$, $h=\{a,b,c\}$, $P(h)=1/3$), we would have similarly looked for a set of $v_a$ obeying $P(a|v_a) \neq 0$, and homogeneous permutations $s$ mapping $v_a$ to $v_b$ (or $v_c$), obeying $P(a|v_a)=P(b|s v_a)$ (or $P(a|v_a)=P(c|s v_a)$). This procedure can in principle be generalized to groups of arbitrary size, we leave however the question of how to do it efficiently to future investigations.

Observe also that focusing on $v \in \partial V_R$ rather than all of $V$ in the above procedure comes at no loss of generality, since as we seek a spatially homogeneous action of the symmetry, the symmetry action on $\partial V_R$ implies its action of $V$. Thus no information is lost and computational resources, related to matching $v_{+}$ and $v_{-}$ are used more efficiently. Had we considered $v \in V$, then, from the perspective of the encoder coupling to the edge $\partial V$, the freedom of flipping $\tau \in V/\partial V$ behaves as a symmetry (with a trivial representation on $H$). However examining the probability $p(v) = \sum_e p(v,e)$ would reveal that it is not a true physical symmetry of $V$.

\section{Numerical Experiments}

To validate our prediction of the the critical temperature $\beta_{c,1}$ and the dependence of the optimal encoder after the IB phase transition on physical quantities we performed a numerical experiment. The test system was the 2D Ising model at criticality. The system was put on a cylinder of three sites' circumference, and the transfer matrix eigenvectors and eigenvalues were obtained by exact diagonalisation to obtain the numerical value of $\beta_{c,1}$ from Eq.\ref{eq:betac1} and the encoder from Eqs.\ref{eq:encoder1} and \ref{eq:mt1}.

This was compared with the numerical solutions to the IB equations presented with the probability distribution of the model configurations. To obtain these solutions we used the simple Iterative IB Algorithm (iIB)\cite{slonim2002information}. The input variables of the iIB algorithm
are: $P(E,V)$, cardinality $|H|$, tradeoff $\beta$, the initial guess for $P(H|V)$, and $\epsilon$, which is a convergence parameter. The outputs are $P(H|V)$, $P(E|H)$ and $P(H)$.

For a given set of input variables, the iIB algorithm iterates between the three IB Eqs.\ref{eq:app_ib_it1}-\ref{eq:app_ib_it3}. On every iteration the first IB equation updates the encoder $P(H|V)$ from the previous iteration. Then, $P(E|H)$ and $P(H)$ are updated according to the remaining equations. The iterations stop when the update on $P(H|V)$ becomes negligible, \emph{i.e.}~if after $n$ iterations $JS[P_{n}(H|V)|P_{n-1}(H|V)]<\epsilon$, where JS is the Jensen-Shannon divergence.  

To produce Fig.\ref{fig:fig3} of the main-text, we applied the iIB algorithm on a range of $\beta$ values. Starting from $\beta=0$ and a maximally symmetric trivial encoder, we increased $\beta$ up to maximal value greater than $\beta_{c,1}$, determined through the numerical analysis of the Hessian (see below). For every $\beta$, in order to prevent the algorithm from getting stuck in a non-optimal local minimum of the Lagrangian, a small random noise was added to the initial guess $P(H|V)$ (the outcome of the optimization for the previous value of $\beta$). The iIB algorithm was then applied on the inputs until convergence. The same steps were then repeated for the next value of $\beta$. 

For this small test system the input $P(E,V)$ distribution was calculated explicitly, using the transfer matrix method. We used a setup where $|V|=2\times 3$,$|E|=1 \times 3$,$L_{B}=9$ and $L=3$. We set the convergence parameter to be $\epsilon=1\mathrm{e}{-14}$, and the random noise added to the encoder was of the order of $1\mathrm{e}{-6}$. 

Just above $\beta_{c,1}$ the iIB algorithm tends to stay around the saddle point given by the trivial uniform $P(H|V)$. This behavior continues until at some higher value of $\beta$ the algorithm converges to the optimal solution of the IB equations through a discontinuity in the IB curve. To fix such numerical artifacts, after each discontinuity we applied the algorithm again, this time by scanning $\beta$ backwards. We initialized the backward scan using the parameters and variables from the forward scan at a $\beta$ value located after the discontinuity point.

To make sure that our backward scan yielded the optimal solution, we compared its IB Lagrangian values with those of the forward scan solution, confirming the former were indeed smaller or equal. Another indication of the instability of the forward scan iIB solution just after $\beta_{c,1}$ was that the eigenvalue of the Hessian which crossed 0 at $\beta_{c,1}$ stayed negative after $\beta_{c,1}$, where the equivalent eigenvalue at the backward scan was positive. 

Finally, we obtain the value of the $146.33999< \beta^{IB}_{c,1}<146.34999$, which agrees with the  $\beta_{c,1}=146.34458$ from the analytical formula Eq.\ref{eq:betac1}, and the encoder in Fig.\ref{fig:fig3}.

We emphasize again, that the analytical results on the IB transition we derived \emph{do not} require the limit of large cylinder circumference $L$, as long as $L_B$ is sufficiently larger than $L$. The prediction is for the IB transition in terms of the transfer matrix eigenvalues and eigenvectors, regardless of the circumference. This is the prediction we verify numerically. In the limit of $L\rightarrow \infty$ the extracted quantities can additionally be related to the CFT scaling operators, per the classical results on transfer matrices and CFTs \cite{CARDY1986186}.

\end{document}